\documentclass[a4paper,11pt]{article}
\usepackage{jheppub}
\usepackage{amsmath}
\usepackage{amssymb}
\usepackage{slashed}
\usepackage{booktabs}
\usepackage{array}
\usepackage[section]{placeins}
\usepackage{needspace}
\newcolumntype{L}[1]{>{\raggedright\arraybackslash}p{#1}}
\newcolumntype{C}[1]{>{\centering\arraybackslash}p{#1}}
\usepackage{xcolor}
\usepackage[compat=1.1.0]{tikz-feynman}
\tikzfeynmanset{warn luatex=false}
\DeclareRobustCommand{\software}[1]{\textnormal{\rmfamily\scshape #1}}
\ifdefined
\fi
\newcommand{\sysname}{\software{FeynFacet}}
\newcommand{\sysnamelogo}{\includegraphics[height=\fontcharht\font`T]{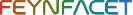}}
\newcommand{\msbar}{\overline{\mathrm{MS}}}
\newcommand{\order}[1]{\mathcal{O}\!\left(#1\right)}
\newcommand{\dd}{\mathrm{d}}
\newcommand{\Lp}{L_\perp}
\newcommand{\Lnu}{L_\nu}
\newcommand{\LQ}{L_Q}
\newcommand{\cI}{\mathcal{I}}
\newcommand{\cC}{\mathcal{C}}
\newcommand{\cD}{\mathcal{D}}
\newcommand{\bT}{b_T}

\title{\texorpdfstring{\sysnamelogo}{FeynFacet}: AI agents for analytic NLO QCD}

\author[a,b,c]{Zhong-Bo Kang,}
\author[d]{Werner Vogelsang,}
\author[a,b]{and Congyue Zhang}
\affiliation[a]{Department of Physics and Astronomy, University of California, Los Angeles, CA 90095, USA}
\affiliation[b]{Mani L. Bhaumik Institute for Theoretical Physics, University of California, Los Angeles, CA 90095, USA}
\affiliation[c]{Center for Frontiers in Nuclear Science, Stony Brook University, Stony Brook, NY 11794, USA}
\affiliation[d]{Institute for Theoretical Physics, University of T\"ubingen, Auf der Morgenstelle 14, D-72076 T\"ubingen, Germany}
\emailAdd{zkang@physics.ucla.edu}
\emailAdd{werner.vogelsang@uni-tuebingen.de}
\emailAdd{maxzhang2002@g.ucla.edu}

\abstract{%
We present \sysname, a workflow in which artificial intelligence agents
develop and execute programs for analytic next-to-leading-order (NLO)
QCD calculations, with the observables and conventions specified by
physicists. We use it to derive new analytic hard functions for
single-inclusive transverse spin transfer in proton--proton and
lepton--proton collisions. {The new proton--proton
results cover all partonic channels contributing to transverse spin
transfer, and the new lepton--proton result is inclusive in the scattered
lepton.} Within \sysname,
we use reverse unitarity to represent the on-shell delta functions in
real-emission phase-space integrals as cut propagators.
{Dimensional shifts} convert integrals containing evanescent
numerators into scalar integrals in higher dimensions.
Integration-by-parts identities then reduce the scalar integrals to
master integrals, which are evaluated analytically.
Gauge invariance, pole cancellation and {renormalization-group (RG) evolution}
provide checks throughout the calculation. We reproduce established
single-inclusive hard functions for unpolarized and double longitudinally polarized proton--proton and lepton--proton scattering, and for proton--proton scattering with two transversely polarized beams. We also reproduce the NLO matching coefficients for unpolarized quark and gluon
transverse-momentum-dependent parton distributions and the associated
soft functions. The new transverse spin-transfer coefficients provide perturbative inputs for extracting quark transversity through spin-dependent fragmentation, including polarized-hyperon production, Collins and dihadron asymmetries in jets, and spin-dependent one- and two-point energy correlators. We make all analytic coefficients and the associated calculation programs publicly available online.}


\begin{document}

\setcounter{tocdepth}{2}
\let\fulltableofcontents\tableofcontents
\renewcommand{\tableofcontents}{\begingroup\renewcommand{\baselinestretch}{0.95}\small\fulltableofcontents\endgroup}
\renewcommand{\afterTocRuleSpace}{\clearpage}
\maketitle

\section{Introduction}
\label{sec:intro}

A quantitative understanding of quantum chromodynamics (QCD) requires
precise scattering predictions and reliable determinations of the
nonperturbative dynamics of hadrons.  These objectives are connected
through QCD factorization~\cite{Collins:1989gx,Collins:2011zzd}.
For observables admitting a factorized
description, cross sections combine perturbatively calculable partonic
coefficients with nonperturbative quantities such as parton distribution
functions (PDFs) and fragmentation functions (FFs), including their
spin-dependent counterparts~\cite{Lin:2017snn,Bertone:2017tyb,
deFlorian:2008mr,Kang:2015msa}.
Perturbative accuracy therefore affects both predictions made with these
functions and their extraction from experimental data.

Higher-order QCD calculations are essential for controlling perturbative
uncertainty and {therefore crucial for} obtaining precise quantitative information from
scattering measurements, including those at the Large Hadron Collider
(LHC), its high-luminosity upgrade, and the future Electron-Ion Collider
(EIC).  At the LHC, accurate QCD predictions underpin precision tests of
the Standard Model and the interpretation of searches for new phenomena.
LHC measurements also constrain the PDFs used in those predictions, so
improvements in the perturbative calculations directly affect the
determination of the partonic structure of the
proton~\cite{Azzi:2019yne,PDF4LHCWorkingGroup:2022cjn}.  At the EIC,
polarized and unpolarized scattering will probe the quark and gluon
structure of nucleons and nuclei, including their spin structure and
multidimensional parton distributions~\cite{AbdulKhalek:2021gbh}.
Extracting this information across the EIC's range of processes and spin
observables requires precise hard functions with the corresponding
kinematic and polarization dependence.

General subtraction methods and
frameworks such as \textsc{MadGraph5\_aMC@NLO} have made numerical
predictions widely accessible~\cite{Frixione:1995ms,Catani:1996vz,Alwall:2014hca}.
Numerical polarized calculations include next-to-leading-order (NLO) jet
production in hadronic~\cite{deFlorian:1998qp} and
lepton--proton~\cite{Boughezal:2018azh} collisions,
{and} single-jet production at next-to-next-to-leading order (NNLO) in
deep-inelastic scattering using projection to
Born~\cite{Borsa:2020ulb,Borsa:2022cap}.
{A \textsc{POWHEG} generator is also available for polarized
proton--proton dijets~\cite{Borsa:2026asi}.}

Analytic higher-order results, by contrast, provide reusable coefficients whose
kinematic dependence, logarithms, and endpoint distributions can be
examined explicitly.  They also connect fixed-order QCD to resummation.
The color and kinematic patterns of the singular terms guide the
construction of a resummed description.  Conversely, expanding a
factorization theorem or an effective theory to a given perturbative
order permits a term-by-term comparison with the corresponding limit
of the full QCD calculation.  This tests the predicted singular
structure, while the finite terms supply the matching between resummed
and fixed-order predictions~\cite{deFlorian:2005yj,Kang:2016mcy,Lee:2024icn}.
The value of an analytic result thus extends beyond a prediction for
one choice of nonperturbative inputs.

{Spin-dependent observables make these advantages particularly important. Global analyses of helicity and transversity
distributions and the fragmentation functions that analyze quark spin
require repeated convolutions with perturbative hard
functions~\cite{deFlorian:2008mr,Nocera:2014gqa,Cocuzza:2022jye,
Kang:2015msa,Cammarota:2020qcw,Gamberg:2022kdb,Zeng:2024gun}.
Polarized distributions of different flavors can have opposite signs,
so cancellations among partonic channels can leave a spin-dependent
cross section much smaller than its individual contributions. Resolving
these cancellations with Monte Carlo integration to the precision
required by a fit can demand high statistics. The polarized hard
functions entering these fits {are defined by differences of partonic cross sections for different spin
configurations}, and their evaluation requires a definite spin prescription
in dimensional regularization: evanescent contributions can multiply
poles and leave finite
terms~\cite{Vogelsang:1996im,Vogelsang:1997ak,Stratmann:2001pt}.
Analytic expressions for each partonic channel improve the numerical
stability of these convolutions and can be reused as the fit inputs
change. They also make the finite contributions from evanescent terms
and any conversion to the chosen parton-distribution scheme explicit.}

Transversity remains the least constrained of the three leading-twist
collinear quark
distributions~\cite{Kang:2015msa,Cammarota:2020qcw,Gamberg:2022kdb}.
It measures the difference
between the densities of quarks polarized parallel and antiparallel to
the spin of a transversely polarized nucleon.  Because it is chiral-odd,
it cannot be measured in inclusive deep-inelastic scattering at leading
twist: it {couples to} a second chiral-odd distribution or fragmentation
function in the cross section.  In single-inclusive transverse spin
transfer, a chiral-odd fragmentation function provides this second
factor, and the hard scattering correlates
the transverse spin of an incoming quark or antiquark with that of the
observed fragmenting quark or antiquark.  Spin-dependent fragmentation
relates this correlation to
hyperon polarization, the Collins azimuthal asymmetry of a hadron inside
a jet, or a dihadron azimuthal
asymmetry~\cite{Bacchetta:2004it,Kang:2017btw,Kang:2020xyq,Kang:2023elg}.
{Near-side energy-energy correlators provide a collinear probe of
transversity through energy-weighted dihadron
fragmentation~\cite{Kang:2026pro,Kang:2026qbm}.}
STAR measurements in polarized-proton collisions at the Relativistic
Heavy Ion Collider (RHIC) already constrain such spin-dependent
correlations~\cite{STAR:2017akg,STARCollaboration:2026kaa}, and the EIC
will extend their study to lepton--proton
scattering~\cite{AbdulKhalek:2021gbh,DAlesio:2026pca}.
Phenomenological studies have combined Born-level production hard functions with evolved distributions and fragmentation functions~\cite{Kang:2017btw}, while NLO matching coefficients for polarized fragmenting jet functions are already available~\cite{Kang:2023elg,Kang:2020xyq}. However, the corresponding NLO production hard functions for the single-inclusive transverse-spin-transfer processes considered here have not previously been available. These missing hard functions are needed to improve the perturbative accuracy of transversity extractions from such measurements.

{Calculating such hard functions analytically involves developing programs
tailored to the process and spin projection.} Agents based on large language models
(LLMs) have recently been used to develop programs for symbolic diagram
calculations, infrared subtraction, and Feynman-integral
evaluation~\cite{Menzo:2026qrl,He:2026lmm,Schwartz:2026BootLoops}.
{Agent-assisted analytic calculations have also been
carried out for vector-boson helicity cross sections at
NLO~\cite{Camarda:2026kdu}.}

We present \sysname,\footnote{{One Feynman-diagram
calculation, many facets: unpolarized and polarized, collinear and
transverse-momentum dependent, NLO and beyond.}} a workflow in which artificial intelligence (AI)
agents develop and execute the process-specific programs for complete
analytic NLO QCD calculations.  Physicists specify the processes,
observables, spin projections, and conventions and assess the physical
interpretation of the results.  The agents reuse shared infrastructure
for amplitude generation, algebra, reduction, and integration, and use
gauge invariance, pole cancellation, and {renormalization-group (RG) evolution} to diagnose and
correct discrepancies in the calculation.

With \sysname, we derive new analytic NLO hard functions for
single-inclusive transverse spin transfer in proton--proton collisions,
covering all seven tagged quark and antiquark classes, and in
lepton--proton collisions with the scattered lepton unobserved.  The
lepton calculation includes the photon-collinear subtraction required
when integrating over the scattered lepton.  {The
{resulting finite coefficients, with the scale dependence determined
by RG evolution, supply the NLO}
partonic production terms for polarized-hyperon production, azimuthal
asymmetries inside jets, and spin-dependent energy
correlators~\cite{Gao:2025evv,Kang:2026qbm}. The corresponding cross
sections follow by convolution with the incoming PDFs and the
appropriate fragmentation or jet functions.}

For the single-inclusive calculations, \sysname\ evaluates the same
spin and color contractions separately with \software{FeynCalc}~\cite{Shtabovenko:2023idz},
\software{FORM}/\allowbreak\software{TFORM}~\cite{Ruijl:2017dtg,Davies:2026cci}
and \software{Tracer}~\cite{Jamin:1991dp}.
We compare their results to check the algebra, then use a common set
of programs for scalar reduction and analytic integration.
{Reverse unitarity~\cite{Anastasiou:2002yz,Anastasiou:2003yy,Anastasiou:2003ds}
expresses real-emission phase-space integrals as cut integrals.
{Dimensional-shift} identities~\cite{Davydychev:1991va,Tarasov:1996br,Bern:1995db}
express integrals containing evanescent numerators as scalar integrals
in higher dimensions.} Integration-by-parts (IBP) identities
derived at a symbolic integration dimension reduce the scalar integrals
for every spin projection in both the original and shifted dimensions.
After analytic integration and
distributional expansion at the endpoint, we combine the real and
virtual contributions with ultraviolet and collinear counterterms to
obtain finite hard functions. Section~\ref{sec:workflow} details the
implementation.

We validate the calculations by reproducing established NLO
single-inclusive hard functions for unpolarized and double-longitudinal
$pp$ and lepton--proton scattering, and for double-transverse
$pp$ scattering~\cite{Aversa:1988vb,Jager:2002xm,Mukherjee:2005rw,Hinderer:2015hra,Hinderer:2017ntk}.
We also reproduce the NLO matching coefficients for unpolarized quark and
gluon transverse-momentum-dependent PDFs (TMDPDFs) and their associated
soft functions~\cite{Luo:2020epw}.  Agreement with these results tests
the treatment of Wilson lines, rapidity divergences, and soft
subtraction.  Table~\ref{tab:calculation-overview} summarizes the
established calculations and the new spin-transfer results.

\begin{table}[htbp]
\centering
\small
\begin{tabular}{@{}>{\centering\arraybackslash}p{0.27\linewidth}>{\centering\arraybackslash}p{0.16\linewidth}>{\centering\arraybackslash}p{0.21\linewidth}>{\centering\arraybackslash}p{0.22\linewidth}@{}}
\toprule
Calculation & Spin & Calculated through & Role in this work\\
\midrule
$pp\to h+X$ & UU, LL & $\alpha_s^3$ & Validation\\[3pt]
$p^\uparrow p^\uparrow\to h+X$ & Incoming TT & $\alpha_s^3$ & Validation\\[3pt]
$\ell p\to h+X$ & UU, LL & $\alpha_{\rm em}^2\alpha_s$ & Validation\\[3pt]
Quark/gluon TMDPDF matching and soft functions & Unpolarized & $\alpha_s$ & Validation\\[3pt]
$p^\uparrow p\to c^\uparrow+X$ & Transfer TT & $\alpha_s^3$ & New hard functions\\[3pt]
$\ell p^\uparrow\to q^\uparrow+X$ & Transfer TT & $\alpha_{\rm em}^2\alpha_s$ & New hard function\\
\bottomrule
\end{tabular}
\caption{Calculations performed with \sysname.  UU and LL denote
unpolarized and double-longitudinal spin configurations.  Incoming TT
has two transversely polarized incoming partons; transfer TT has one
transversely polarized incoming parton and a transversely polarized
observed outgoing parton.  In the collinear validation rows, $h$ is
unpolarized.  In the transfer TT rows, $c$ and $q$ denote the tagged quark
or antiquark.  The scattered
lepton is unobserved in all lepton processes, and the listed
$\alpha_{\rm em}^2\alpha_s$ accuracy refers to the direct-lepton hard
functions.  The direct- and resolved-photon Born hard functions are also
computed for the lepton--proton applications.}
\label{tab:calculation-overview}
\end{table}

The analytic coefficients and calculation programs will be publicly
released with this paper at \url{https://github.com/qcdphysics/FeynFacet}.
The accompanying project webpage at
\url{https://qcdphysics.github.io/FeynFacet/} will provide further
documentation on the processes and workflows presented in this paper, along with an example prompt illustrating how to use Codex with FeynFacet to perform new calculations.

The rest of the paper is organized as follows. Section~\ref{sec:physics} defines the observables, spin projections, and
conventions.  Section~\ref{sec:workflow} presents the calculation and its
physics checks.  Section~\ref{sec:benchmarks} presents benchmarks
against established collinear hard functions, TMD matching coefficients and soft functions.
Section~\ref{sec:newresults} presents the new spin-transfer
hard functions and their applications.  The outlook discusses
higher-order extensions, and the appendices give the coefficient and
computational details.

\section{Theoretical framework}
\label{sec:physics}

\subsection{Observables, collinear factorization, and hard functions}
\label{sec:observables}
\label{sec:conventions}
\label{sec:distributions}

{We consider single-inclusive hadron and jet production
at large transverse momentum in lepton--proton and proton--proton
collisions. We denote the measured object by $\mathcal O=h$ or
$\mathrm{jet}$, its momentum by $P_{\mathcal O}$, and its transverse
momentum by $p_T$. In collinear factorization, the hard functions
describe production of the parton that fragments into the hadron or
initiates the jet.}
All partonic hard functions use the normalization
\begin{equation}
 \widehat\sigma(s,t,u,\mu)
 \equiv sE_p\frac{\dd^3\widehat\sigma_{\rm part}}{\dd^3p},
\end{equation}
where $p$ is the tagged parton momentum and $E_p$ its energy. The scale
argument $\mu$
collectively denotes the renormalization and factorization scales
specified below.

{The lepton--proton processes are
\begin{equation}
 \ell(l)+p(P)\longrightarrow h(P_h)+X,\qquad
 \ell(l)+p(P)\longrightarrow \mathrm{jet}(P_J)+X,
 \label{eq:lN-process}
\end{equation}
with the scattered lepton included in the unobserved state $X$~\cite{Kang:2011jw,Gamberg:2014eia}.
The transverse momentum $p_T$ sets the hard
scale~\cite{Hinderer:2015hra,Hinderer:2017ntk,Abelof:2016pby,Boughezal:2018azh}. {With masses neglected at leading power}, the hadronic invariants are}
\begin{equation}
 S=(P+l)^2,\qquad T=(P-P_{\mathcal O})^2,\qquad
 U=(l-P_{\mathcal O})^2.
\end{equation}
The incoming and outgoing parton momenta are $k=xP$ and
$p=P_{\mathcal O}/z$, respectively, where $x$ is the proton momentum
fraction and $z$ the fraction of the outgoing parton's momentum carried
by $\mathcal O$. The partonic invariants are $s=(k+l)^2=xS$,
$t=(k-p)^2=xT/z$, and $u=(l-p)^2=U/z$. In the massless-lepton
formulation,
the direct-lepton and direct-photon contributions
are{~\cite{Hinderer:2015hra,Hinderer:2017ntk}}
\begin{align}
 E_{\mathcal O}\frac{\dd^3\sigma^{\ell p\to\mathcal O X}}
 {\dd^3P_{\mathcal O}}
 ={}&\frac{1}{S}\sum_{i,f}\int_0^1\frac{\dd x}{x}
 \int_0^1\frac{\dd z}{z^2}\,
 f^{i/p}(x,\mu)\,{\cal F}^{\mathcal O}_f(z,\mu)
 \Bigg[\widehat\sigma^{\,\ell i\to f}(s,t,u,\mu)\nonumber\\
 &\quad+\int_0^1\frac{\dd y}{y}\,
 f_{\gamma/\ell}(y,\mu_L)
 \widehat\sigma^{\,\gamma i\to f}(ys,t,yu,\mu)\Bigg].
 \label{eq:lN-factorization}
\end{align}
Here $f^{i/p}$ is the PDF for the incoming parton $i$ in the proton, ${\cal F}^{\mathcal O}_f$ is the
final-state factor defined below, and $\widehat\sigma^{\,\ell i\to f}$
denotes the direct hard function for $\ell+i\to f+X$.
The photon distribution $f_{\gamma/\ell}$ absorbs the lepton-collinear
singularity at the scale $\mu_L$ and contains the resulting logarithmic
dependence on the physical lepton mass. The photon hard function has the same invariant
normalization and is evaluated for incoming photon momentum $yl$.

The direct-lepton Born process is $\ell q\to\ell q$. Through NLO,
the contributing channels are $\ell q\to qX$, $\ell q\to gX$, and
$\ell g\to qX$, together with the corresponding antiquark channels.
The explicitly displayed final-state parton fragments into the observed
hadron, while $X$ contains the scattered lepton and all other unobserved
particles.
The photon Born channels are $\gamma q\to qX$, $\gamma q\to gX$,
and $\gamma g\to qX$, together with their charge conjugates.
The unpolarized and helicity-dependent photon-in-lepton splitting
kernels are
\begin{equation}
P_{\gamma/\ell}(y)=\frac{1+(1-y)^2}{y},\qquad
 \Delta P_{\gamma/\ell}(y)=2-y,
\end{equation}
with the common coupling factor $\alpha_{\rm em}/(2\pi)$
removed~\cite{Hinderer:2015hra,Hinderer:2017ntk}.

In direct photoproduction, the photon itself enters the hard
subprocess. In resolved photoproduction, a quark, antiquark or gluon
from the photon enters the hard subprocess. The distribution
$f_{b/\gamma}(r,M_\gamma)$ describes parton $b$ carrying a fraction $r$
of the photon momentum. The resolved-photon contribution used in
Sec.~\ref{sec:pheno} adds
\begin{equation}
 \sum_b\int_0^1\frac{\dd y}{y}\int_0^1\frac{\dd r}{r}\,
 f_{\gamma/\ell}(y,\mu_L)f_{b/\gamma}(r,M_\gamma)
 \widehat\sigma^{\,ib\to f}(rys,t,ryu,\mu)
\end{equation}
to the square bracket in Eq.~\eqref{eq:lN-factorization}.
The nucleon parton is listed first, so $t$ remains the invariant
between that parton and the tagged parton. The photon PDFs
$f_{b/\gamma}$ have a separate factorization scale $M_\gamma$.
The direct-lepton hard functions are evaluated at NLO, while direct
and resolved photoproduction enter at Born accuracy, as detailed
in Sec.~\ref{sec:lp-uu}.

{The proton--proton
processes~\cite{Aversa:1988vb,Jager:2002xm,Kang:2016mcy} are
\begin{equation}
 p(P_a)+p(P_b)\longrightarrow h(P_h)+X,\qquad
 p(P_a)+p(P_b)\longrightarrow \mathrm{jet}(P_J)+X,
 \label{eq:pp-process}
\end{equation}
with hadronic invariants}
\begin{equation}
 S=(P_a+P_b)^2,\qquad T=(P_a-P_{\mathcal O})^2,\qquad
 U=(P_b-P_{\mathcal O})^2.
\end{equation}
With $k_a=x_aP_a$, $k_b=x_bP_b$, and $p=P_{\mathcal O}/z$,
where $x_a,x_b$ are the incoming proton momentum fractions, the partonic
invariants are $s=(k_a+k_b)^2=x_ax_bS$, $t=(k_a-p)^2=x_aT/z$,
and $u=(k_b-p)^2=x_bU/z$.
The factorized cross section is~\cite{Aversa:1988vb,Jager:2002xm,Kang:2016mcy,Kaufmann:2015hma,Dai:2016hzf}
\begin{align}
 E_{\mathcal O}\frac{\dd^3\sigma^{pp\to\mathcal O X}}{\dd^3P_{\mathcal O}}
 ={}&\frac{1}{S}\sum_{a,b,c}\int_0^1\frac{\dd x_a}{x_a}
 \int_0^1\frac{\dd x_b}{x_b}\int_0^1\frac{\dd z}{z^2}\,
 f^{a/p}(x_a,\mu)f^{b/p}(x_b,\mu)\nonumber\\
 &\hspace{12mm}\times{\cal F}^{\mathcal O}_c(z,\mu)
 \widehat\sigma^{\,ab\to c}(s,t,u,\mu).
 \label{eq:pp-factorization}
\end{align}
The hard functions vanish outside physical partonic phase space,
which fixes the support of the displayed convolutions. The contributing
partonic channels and tagged-parton classes are listed in
Sec.~\ref{sec:pp-uu}.

{The measured final state enters the factorization
formulas through}
\begin{equation}
 {\cal F}^{h}_c(z,\mu)=D^{h/c}(z,\mu),\qquad
 {\cal F}^{\mathrm{jet}}_c(z,\mu)={\cal J}_c(z,p_T R,\mu),
 \label{eq:hadron-jet-replacement}
\end{equation}
{where $D^{h/c}$ is the fragmentation function and ${\cal J}_c$ is the so-called semi-inclusive jet function~\cite{Kang:2016mcy,Kaufmann:2015hma,Dai:2016hzf} for a jet carrying fraction $z$ of the parent
parton momentum. The jet function depends on the jet
algorithm and radius $R$. We use its leading-power form for $R\ll1$,
with a perturbative jet scale $p_T R\gg\Lambda_{\rm QCD}$.}

{At fixed measured jet momentum $p_T$, the parent parton
transverse momentum is $p_{T,c}=p_T/z$. Writing the jet function as
$J_c(z,L,\mu)$, with $L=\ln[\mu^2/(p_{T,c}R)^2]$, gives
\begin{equation}
 {\cal J}_c(z,p_T R,\mu)
 =J_c\!\left(z,L_J+2\ln z,\mu\right),\qquad
 L_J=\ln\frac{\mu^2}{p_T^2R^2}.
 \label{eq:jet-parent-convention}
\end{equation}
The final-state integral thus samples both the momentum-fraction
dependence and the logarithmic scale dependence of $J_c$ as $z$ varies.
{This modified convolution for inclusive jet production, derived
in Ref.~\cite{Lee:2024icn}, improves on the convolution in
Refs.~\cite{Kang:2016mcy,Kang:2016ehg}.}}

{For measurements of hadrons or hadron pairs inside a
jet, the final-state factor is the corresponding fragmenting jet
function, differential in the measured substructure
variables~\cite{Kang:2016ehg,Kang:2017glf,Kang:2023elg}.
Section~\ref{sec:transfer-observables} develops the spin-dependent
applications.}

The hard functions are finite after ultraviolet renormalization and
initial- and final-state collinear subtraction in the modified
minimal-subtraction ($\msbar$) scheme. We write
$\widehat\sigma=\widehat\sigma^{(0)}+\widehat\sigma^{(1)}+\cdots$,
including the couplings in each term; the NLO correction carries one
additional power of $\alpha_s$ relative to the Born process.
{With the counterterm signs included in their definitions},
\begin{equation}
 \widehat\sigma_{\rm finite}^{(1)}
 =\lim_{\epsilon\to0}\left[
 \widehat\sigma_{\rm real}^{(1)}+
 \widehat\sigma_{\rm virtual}^{(1)}+
 \widehat\sigma_{\rm UV}^{(1)}+
 \widehat\sigma_{\rm coll}^{(1)}\right].
 \label{eq:finite-nlo-assembly}
\end{equation}
Ultraviolet counterterms remove the short-distance poles, soft poles
cancel between real and virtual contributions, and collinear poles
are absorbed into the parton distributions and final-state functions.
The collinear counterterm $\widehat\sigma_{\rm coll}^{(1)}$ includes
the initial- and final-state subtractions. These convolve splitting kernels with Born
hard functions computed in $D=4-2\epsilon$ dimensions. {The $\epsilon$-dependent terms of these Born hard functions contribute to the finite NLO result}: a contribution
$(B_0+\epsilon B_1+\cdots)/\epsilon$ contains the finite term $B_1$.

For each hard subprocess, we parametrize the dimensionless kinematics by
\begin{equation}
 v=1+\frac{t}{s},\qquad w=-\frac{u}{s+t},
 \label{eq:vw}
\end{equation}
so that the invariant mass squared of the unobserved recoil is
$s_4=s+t+u=sv(1-w)$. In the physical region $s>0$ and $0<v,w<1$;
two-body kinematics lies at $w=1$. Born terms and virtual corrections
are proportional to $\delta(1-w)$, while real emission also contributes
for $w<1$. Suppressing the dependence on $s$ and the scales, we express
the finite NLO result as
\begin{align}
 \widehat\sigma^{(1)}(v,w)
 ={}&\mathcal N(v,w)R(v,w)\nonumber\\
 &+\mathcal N(v,1)\Big\{A(v)\delta(1-w)+B(v)\cD_0(w)
                         +C(v)\cD_1(w)\Big\},
 \label{eq:distribution-form}
\end{align}
with plus distributions on $0\leq w\leq1$ defined by
\begin{equation}
 \cD_n(w)\equiv\left[\frac{\ln^n(1-w)}{1-w}\right]_+.
 \label{eq:plus-distributions}
\end{equation}
The normalization $\mathcal N$ contains couplings, color factors and
kinematic prefactors. Its endpoint value multiplies the distributions;
the regular remainders generated by this rearrangement are included in
$R$. This is the convention used in the coefficient tables. The regular
term has no plus prescription and may contain integrable logarithms
such as $\ln(1-w)$.

The endpoint distributions follow from
\begin{equation}
 (1-w)^{-1-a\epsilon}
 =-\frac{\delta(1-w)}{a\epsilon}+\cD_0(w)
     -a\epsilon\cD_1(w)+\order{\epsilon^2},
 \label{eq:endpoint-expansion}
\end{equation}
acting on test functions smooth at $w=1$. The coefficient $A$ receives
the virtual contribution, the real-emission endpoint and the delta
terms of the counterterms. The coefficients $B$ and $C$ multiply the
threshold distributions generated by soft and collinear radiation.

We distinguish the strong-coupling renormalization scale $\mu_R$,
the incoming-PDF scale $\mu_F$, the final-state factorization scale
$\mu_F'$, and the photon-in-lepton factorization scale $\mu_L$.
The direct-lepton UU and LL tables use
$\mu_R=\mu_F=\mu_F'=\mu_L=\mu$.
Table~\ref{tab:scales} collects the scale notation. The color factors
are $C_A=N_c$, $C_F=(N_c^2-1)/(2N_c)$ and $T_F=1/2$;
$N_f$ is the number of active massless quark flavors.
The couplings satisfy $e^2=4\pi\alpha_{\rm em}$ and
$g_s^2=4\pi\alpha_s$, and $e_q$ denotes the quark electric charge
in positron units.

\begin{table}[htbp]
 \centering\small
 \begin{tabular}{cL{.66\linewidth}}
 \toprule
 Scale or logarithm & Role\\
 \midrule
 $\mu_R$, $L_R=\ln(\mu_R^2/s)$ & Strong-coupling renormalization\\[3pt]
 $\mu_F,\mu_F'$ & Incoming-PDF and final-state factorization;
 $L_F=\ln(\mu_F^2/s)$, $L_F'=\ln[(\mu_F')^2/s]$\\[3pt]
 $\mu_L$ & Photon-in-lepton factorization;
 $L_L=\ln(\mu_L^2/s)$\\[3pt]
 $M_\gamma$ & Parton-in-photon factorization in the resolved term\\[3pt]
 $\mu$, $L_J$ & Jet matching at $p_T R$;
 $L_J=\ln[\mu^2/(p_T R)^2]$\\[3pt]
 $\mu,\nu,\zeta$ & TMD ultraviolet, auxiliary rapidity and
 Collins--Soper scales, respectively\\
 \bottomrule
 \end{tabular}
 \caption{Scale conventions. The jet logarithm $L_J$ uses the measured
 jet momentum and is distinct from the renormalization logarithm $L_R$.}
 \label{tab:scales}
\end{table}

\FloatBarrier
\subsection{Polarized cross sections and the HVBM scheme}
\label{sec:spin}

The leading-twist collinear quark distributions in a spin-$\tfrac12$
nucleon are the unpolarized distribution $f_1$, the helicity distribution
$g_1$, and the transversity distribution $h_1$. The latter two measure
the differences between quark densities with spin aligned and
antialigned {with the spin of a longitudinally or transversely polarized
nucleon, respectively}.
The gluon has unpolarized and helicity distributions; a spin-$\tfrac12$
target has no leading-twist collinear gluon transversity~\cite{Jaffe:1989xy}. Since $h_1$
is chiral odd, it enters the cross section with a second transversity
distribution or a chiral-odd {transverse spin analyzer}.

Table~\ref{tab:spin-conventions} defines the partonic spin averages and
differences. The same initial-spin combinations define the hadronic
$\dd\sigma_{LL}$ and $\dd\sigma_{TT}$~\cite{Jager:2002xm,Mukherjee:2005rw}.
For transverse spin transfer, the full outgoing-spin difference defines
the hard function $\widehat\sigma_T$. The {transverse spin
analyzer} ${\cal A}^c$ describes the dependence of the measured hadron
polarization or azimuthal distribution on the outgoing parton's
transverse spin. Its convolution with the hard function and incoming
PDFs gives the structure function $F_T[{\cal A}]$ in
Eq.~\eqref{eq:pp-transfer-analyzer} {below}. Examples
include polarized-hyperon fragmentation, Collins hadron-in-jet
production and dihadron interference fragmentation, discussed in
Sec.~\ref{sec:transfer-observables}.

\begin{table}[htbp]
 \centering\footnotesize
 \setlength{\tabcolsep}{4pt}
 \begin{tabular}{@{}L{.13\linewidth}L{.54\linewidth}L{.27\linewidth}@{}}
 \toprule
 Projection & Partonic hard function & Factors in the $pp$ convolution\\
 \midrule
 UU & $\widehat\sigma$: incoming spin average, final spin sum
 & $f_1^{a/p}f_1^{b/p}D_1^{h/c}$\\[5pt]
 LL & $\widehat\sigma_{LL}=\tfrac12[\widehat\sigma(++)-
 \widehat\sigma(+-)]$
 & $g_1^{a/p}g_1^{b/p}D_1^{h/c}$\\[5pt]
 Incoming TT & $\widehat\sigma_{TT}=\tfrac12[
 \widehat\sigma(\uparrow\uparrow)-
 \widehat\sigma(\uparrow\downarrow)]$
 & $h_1^{a/p}h_1^{b/p}D_1^{h/c}$\\[5pt]
 Transfer TT & $\widehat\sigma_T=
 \widehat\sigma(\uparrow\uparrow')-
 \widehat\sigma(\uparrow\downarrow')$
 & $h_1^{a/p}f_1^{b/p}{\cal A}^{c}$\\
 \bottomrule
 \end{tabular}
 \caption{Spin conventions for $ab\to c+X$ and the corresponding
 distribution products in proton--proton collisions, with flavor sums
 and convolutions suppressed. The first three rows describe
 $pp\to hX$ with unpolarized fragmentation, $D_1^{h/c}=D^{h/c}$.
 In LL and incoming TT, the signs and arrows refer to the incoming spins. In the
 transfer TT row, $a,c=q,\bar q$, parton $b$ is unpolarized, and the
 primed arrow denotes the outgoing spin along $\pm s_T'$.}
 \label{tab:spin-conventions}
\end{table}

For direct lepton scattering the corresponding distribution factors
are $f_1D_1$ in UU, $g_1D_1$ in LL, and $h_1{\cal A}$ for transfer.
LL requires a longitudinally polarized lepton; the lepton is
unpolarized in the transfer calculation. Consequently, the
photon contribution uses $\Delta f_{\gamma/\ell}$ for LL and
$f_{\gamma/\ell}$ for transverse spin transfer~\cite{Hinderer:2017ntk}.

The spin sums and differences are evaluated with fermion density
matrices and gauge-boson polarization projectors. For an incoming
massless quark of momentum $k$, the unpolarized density matrix is
$\slashed k/2$; summing over the spin of an outgoing quark of momentum
$p$ gives $\slashed p$.
{For each unpolarized incoming gluon, the physical
polarization sum is multiplied by $1/(D-2)$. Final-state gluon
polarizations are summed.}
For LL, helicity-difference projectors are inserted on both incoming
legs. The quark insertion is $\slashed k\gamma_5/2$ in our convention,
with the corresponding charge-conjugated matrix for an antiquark.

For incoming TT, both polarized quark legs use the insertion
$\slashed k\gamma_5\slashed s_T/2$, where $s_T$ is the unit transverse
spin four-vector of the incoming quark, with $k\cdot s_T=0$ and
$s_T^2=-1$. For transverse spin transfer, the same incoming insertion
is paired with the outgoing insertion $\slashed p\gamma_5\slashed s_T'$,
where $s_T'$ is the outgoing quark's unit transverse spin four-vector,
with $p\cdot s_T'=0$ and $(s_T')^2=-1$.
The outgoing insertion gives the full spin difference in
Table~\ref{tab:spin-conventions}; a hard function defined with the
outgoing half-difference is $\widehat\sigma_T/2$.

In the partonic center-of-mass frame, $N$ is the common normal to the
scattering plane and $X$ lies in that plane and is transverse to the
corresponding quark momentum. The outgoing $X$ axis points toward
increasing polar angle. The $\cos(2\Phi)$ projection for incoming TT,
where $\Phi$ is the azimuth of the tagged parton relative to the common
transverse-spin axis, is specified in Sec.~\ref{sec:pp-tt}.
The relation to incoming TT and the Collins
projection are given in Sec.~\ref{sec:pp-transfer} and
Eq.~\eqref{eq:collins-tensor}.

We use the 't~Hooft--Veltman--Breitenlohner--Maison (HVBM)
prescription~\cite{tHooft:1972tcz,Breitenlohner:1977hr} for dimensional
regularization with $D=4-2\epsilon$. The metric and
Dirac matrices are decomposed into physical four-dimensional and
evanescent parts,
\begin{equation}
 g^{\mu\nu}=\bar g^{\mu\nu}+\widehat g^{\mu\nu},\qquad
 \bar g^\mu{}_{\mu}=4,\qquad
 \widehat g^\mu{}_{\mu}=D-4=-2\epsilon,\qquad
 \bar g^{\mu\nu}\widehat g_{\nu\rho}=0,
 \label{eq:hvbm-metric}
\end{equation}
with
\begin{equation}
 \{\gamma_5,\bar\gamma^\mu\}=0,\qquad
 [\gamma_5,\widehat\gamma^\mu]=0,\qquad
 \gamma_5^2=1.
 \label{eq:hvbm-gamma5}
\end{equation}
The Levi-Civita tensor, incoming and observed momenta, and physical
spin vectors lie in the four-dimensional subspace.
{We use
$\operatorname{tr}(\bar\gamma^\mu\bar\gamma^\nu\bar\gamma^\rho
\bar\gamma^\sigma\gamma_5)=-4i\epsilon^{\mu\nu\rho\sigma}$.}
{For an incoming gluon of momentum $k$, the LL
helicity-difference insertion is
$i\epsilon^{\mu\nu\alpha\beta}k_\alpha n_\beta/(2k\cdot n)$,
with $n$ a lightlike reference vector satisfying $k\cdot n\ne0$.
The factor $1/2$ follows the helicity-difference normalization in
Table~\ref{tab:spin-conventions}.} Loop and
unobserved real-emission momenta, together with sums over unmeasured
polarizations, are treated in $D$ dimensions. Scalar products in the
evanescent subspace are retained through integration, since the
resulting dimensional factors can multiply poles and leave finite
contributions. Traces containing two $\gamma_5$ matrices, including
those in the transfer calculation, are evaluated with
Eq.~\eqref{eq:hvbm-gamma5}.

In HVBM, the dimensionally continued quark splitting functions obey
\begin{equation}
 \Delta P_{qq}^{(4-2\epsilon)}(z)-P_{qq}^{(4-2\epsilon)}(z)
 =4C_F\,\epsilon\,(1-z),\qquad
 \Delta f_{qq}(z)=-4C_F(1-z),
 \label{eq:hvbm-restoration}
\end{equation}
in the conventions of Refs.~\cite{Jager:2002xm,Hinderer:2017ntk}.
The evanescent difference leaves a finite term when multiplied by
the collinear pole. To restore the helicity convention of the
polarized $\msbar$ PDFs, the hard function receives
$-(\alpha_s/2\pi)\Delta f_{qq}\otimes
\widehat\sigma_{LL}^{(0)}$ on each longitudinally polarized
incoming quark or antiquark leg~\cite{Vogelsang:1996im}.
In the pure-annihilation channel $q_i\bar q_i\to q_k+X$, $k\ne i$,
massless helicity conservation then gives
$\widehat\sigma_{LL}=-\widehat\sigma$~\cite{Jager:2002xm}.

The $\msbar$ collinear counterterms on each transversely polarized
incoming or fragmenting quark or antiquark leg use
$P^T_{qq}(z)=C_F[2z/(1-z)_++\tfrac32\delta(1-z)]$,
common to space-like and time-like evolution at leading
order~\cite{Vogelsang:1997ak,Stratmann:2001pt,Mukherjee:2005rw}.
For these counterterms, the unresolved transverse momentum is
averaged over its $D-2$ transverse directions while the physical
spin vectors remain four-dimensional. After this angular average,
the terms linear in $\epsilon$ cancel in the real transversity
kernel~\cite{Vogelsang:1997ak},
\begin{equation}
 P^{T,\mathrm{real}}_{qq}(z,\epsilon)
 =C_F\frac{2z}{1-z}+\order{\epsilon^2},
 \qquad 0<z<1.
 \label{eq:transverse-D-kernel}
\end{equation}

\FloatBarrier
\subsection{TMD matching coefficients and the soft function}
\label{sec:tmd-setup}

We also compute the NLO matching coefficients of unpolarized quark
and gluon TMDPDFs and the associated soft functions. We follow the
operator and rapidity-subtraction conventions of
Refs.~\cite{Luo:2019hmp,Luo:2020epw} and compare our results with
theirs in Secs.~\ref{sec:tmd-soft} and~\ref{sec:tmd}.

Choose lightlike directions $n,\bar n$ with $n\cdot\bar n=2$ and
write $k^+=n\cdot k$, $k^-=\bar n\cdot k$. The target has large
light-cone momentum $P^+=n\cdot P$, and a massless cut momentum
satisfies $k^+k^-=\boldsymbol k_T^2$. The transverse separation is
$\bT=|\boldsymbol b_T|$. We denote the longitudinal momentum entering
the TMD operator by $Q=xP^+$ and use the Collins--Soper scale
$\zeta=Q^2$. The momentum $Q$ is held fixed in convolutions over the
matching fraction $\xi$.

For an unpolarized target, the unsubtracted beam functions are
\begin{align}
 B^{\rm unsub}_{q/N}(x,\bT)&=\frac12\int\frac{\dd\lambda}{2\pi}\,
 e^{-ixP^+\lambda}\,
 \langle N(P)|O_q(\lambda n+\boldsymbol b_T,0)|N(P)\rangle,\notag\\
 B^{\rm unsub}_{g/N}(x,\bT)&=\frac{1}{xP^+}\int\frac{\dd\lambda}{2\pi}\,
 e^{-ixP^+\lambda}\,
 \langle N(P)|O_g(\lambda n+\boldsymbol b_T,0)|N(P)\rangle,
 \label{eq:tmd-beam-definition}
\end{align}
with bare operators
\begin{align}
 O_q(y,0)&=\bar q(y)U_{n,F}(y,0)\slashed n\,q(0),\notag\\
 O_g(y,0)&=-g_{\perp\mu\nu}\,n_\rho F^{a,\rho\mu}(y)
 U_{n,A}^{ab}(y,0)n_\sigma F^{b,\sigma\nu}(0),
 \label{eq:tmd-bilocal-operators}
\end{align}
where $g_\perp^{\mu\nu}=g^{\mu\nu}
-(n^\mu\bar n^\nu+\bar n^\mu n^\nu)/2$.
The staple-shaped Wilson line $U_{n,R}(y,0)$ transports color from
$0$ to $y=\lambda n+\boldsymbol b_T$ through future light-cone
infinity in representation $R$, including the transverse link at
infinity. The unpolarized functions considered here are unchanged
for past-pointing lines. The transverse Fourier convention is
$B(x,\boldsymbol b_T)=\int\dd^{D-2}\boldsymbol k_T\,
e^{i\boldsymbol b_T\cdot\boldsymbol k_T}B(x,\boldsymbol k_T)$.

Dimensional regularization does not regulate the divergences from
arbitrarily large emission rapidities at fixed transverse momentum.
We use the exponential regulator~\cite{Li:2016axz}, weighting a real
state of total momentum $k$ by
$\exp[-\tau e^{-\gamma_E}(k^++k^-)]$ in both beam and soft matrix
elements. After integration, we retain the leading small-$\tau$ terms,
including logarithms of $\tau$, at fixed $\epsilon$ before expanding
in $\epsilon$. The auxiliary rapidity scale is
$\nu=1/\tau$~\cite{Luo:2019hmp}.

Let $Y_{v,R}(x)$ be a soft Wilson line transporting color from future
infinity to $x$ along $v$, and define
$J_R=Y_{n,R}^\dagger Y_{\bar n,R}$. The regulated bare soft function is
\begin{align}
 S_R^{\rm bare}(\bT,\tau)=\frac{1}{d_R}\sum_X
 e^{i\boldsymbol b_T\cdot\boldsymbol k_{XT}}
 e^{-\tau e^{-\gamma_E}(k_X^++k_X^-)}
 \operatorname{Tr}\!\big[&
 \langle0|\overline{\mathrm T}\{J_R^\dagger(0)\}|X\rangle\notag\\[-2pt]
 &\times\langle X|\mathrm T\{J_R(0)\}|0\rangle\big],
 \label{eq:tmd-soft-cut-operator}
\end{align}
where the sum includes phase space, $k_X$ is the total real momentum,
and $\mathrm T$ and $\overline{\mathrm T}$ denote time and anti-time
ordering. The color trace is normalized by $d_F=N_c$ or
$d_A=N_c^2-1$, so that $S_R^{(0)}=1$.
The renormalized soft functions are $S_q=S_F$ and $S_g=S_A$ after
$\msbar$ subtraction. The soft overlap (zero-bin) in the unsubtracted beam
function is removed by division by the corresponding bare soft
function. Beam-operator renormalization in $\msbar$ then gives
$B_{i/N}(x,\bT,\mu,\nu;Q)$.

For $\bT\Lambda_{\rm QCD}\ll1$, this beam function admits the
operator-product expansion
\begin{equation}
 B_{i/N}(x,\bT,\mu,\nu;Q)
 =\sum_j\int_x^1\frac{\dd\xi}{\xi}\,
 \cI_{ij}(\xi,\bT,\mu,\nu;Q)f^{j/N}(x/\xi,\mu)
 +\order{\bT^2\Lambda_{\rm QCD}^2}.
 \label{eq:tmd-ope}
\end{equation}
The matching coefficients are extracted with spin- and color-averaged
partonic external states by matching onto the renormalized collinear
PDFs. In a term of this convolution the parent parton has momentum
$p^+=(x/\xi)P^+$, so that $\xi p^+=Q$. The normalization in
Eq.~\eqref{eq:tmd-beam-definition} gives
$B^{(0)}_{i/j}=\delta_{ij}\delta(1-\xi)$.

The soft-subtracted hadronic TMDPDF~\cite{Boussarie:2023izj} and its small-$\bT$ expansion are
\begin{align}
 F_{i/N}(x,\bT,\mu,Q^2)
 &=B_{i/N}(x,\bT,\mu,\nu;Q)\sqrt{S_i(\bT,\mu,\nu)}\notag\\
 &=\sum_j\int_x^1\frac{\dd\xi}{\xi}\,
 \cC_{ij}(\xi,\bT,\mu,Q^2)f^{j/N}(x/\xi,\mu)
 +\order{\bT^2\Lambda_{\rm QCD}^2},
 \label{eq:tmd-hadronic-distribution}
\end{align}
with
\begin{equation}
 \cC_{ij}(\xi,\bT,\mu,Q^2)
 =\cI_{ij}(\xi,\bT,\mu,\nu;Q)\sqrt{S_i(\bT,\mu,\nu)}.
 \label{eq:tmd-physical}
\end{equation}
The dependence on $\nu$ cancels in this product at fixed $Q$, while
the dependence on $\mu$ and $\zeta=Q^2$ remains.

We express the results in terms of
\begin{equation}
 a_s=\frac{\alpha_s(\mu)}{4\pi},\qquad
 \Lp=\ln\frac{\bT^2\mu^2}{b_0^2},\qquad b_0=2e^{-\gamma_E},\qquad
 \LQ=2\ln\frac{Q}{\nu}=\ln\frac{\zeta}{\nu^2},
 \label{eq:tmd-logs}
\end{equation}
and $\Lnu=\ln(\nu^2/\mu^2)$, with expansions
$\cI_{ij}=\sum_m a_s^m\cI^{(m)}_{ij}$ and similarly for $S_i$
and $\cC_{ij}$. At LO the diagonal beam coefficients are
$\cI^{(0)}_{qq}=\cI^{(0)}_{gg}=\delta(1-\xi)$ and the
off-diagonal coefficients vanish. In the combined NLO coefficients,
the rapidity logarithms occur through
$\LQ+\Lnu=\ln(\zeta/\mu^2)$.

\section{The \sysname\ workflow}
\label{sec:workflow}
\label{sec:routes}
\label{sec:principle}

\sysname\ connects spin-projected partonic amplitudes to the finite NLO
hard functions defined in Sec.~\ref{sec:observables}. We describe the
single-inclusive collinear calculation, following the real and virtual
contributions through scalar reduction, analytic integration and collinear
subtraction. The TMD application uses the same algebraic and
integration methods for the operator matrix elements in
Secs.~\ref{sec:tmd-soft} and~\ref{sec:tmd}; the measured integrals are
given in Appendix~\ref{app:tmd-integrals}.

The process, observable, spin projection and subtraction scheme are
specified by the physicists. Under these requirements, AI
agents reuse the process-independent infrastructure and develop and execute the process-specific programs that implement
the calculation. The completed programs run without further
language-model calls. Figure~\ref{fig:workflow} summarizes the
calculation; the development procedure is described in
Appendix~\ref{app:sessions}.

\begin{figure}[t]
 \centering
 \includegraphics[width=\textwidth]{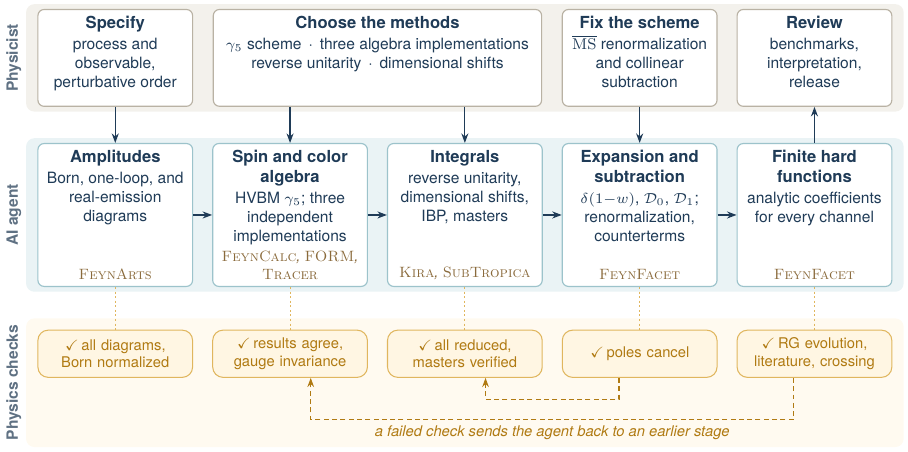}
 \caption{The \sysname\ calculation of single-inclusive NLO hard
 functions. The physicist specifies the observable, methods and
 conventions, and the agent develops and executes the programs for the
 five computational stages. The amplitudes, scalar integrals and
 subtraction terms are checked at the indicated stages. Dashed arrows
 indicate feedback during development. Figure~\ref{fig:workflow-chart} in
 Appendix~\ref{app:spec} gives a more detailed calculation chart.}
 \label{fig:workflow}
\end{figure}

\subsection{Amplitudes and spin algebra}

We use \software{FeynArts}~\cite{Hahn:2000kx} to generate the Born, one-loop
and real-emission amplitudes, together with the lower-order amplitudes
needed for the collinear counterterms.\footnote{{An
alternative is \software{QGRAF}~\cite{Nogueira:1991ex}, whose output can
be converted into \software{FeynCalc} amplitudes through the
\software{FeynHelpers}~2 interface~\cite{Shtabovenko:2025lxq}.}}
Amplitudes for single-inclusive
parton production are generated with the existing \software{FeynArts} models. The model used for the TMD
calculation includes the Wilson-line interactions and operator endpoints
described in Appendix~\ref{sec:tmd-model}. From these amplitudes we form the
Born and real-emission squared matrix elements and the virtual interference
$2\operatorname{Re}[\mathcal{M}^{(1)}\mathcal{M}^{(0)*}]$, with
$\mathcal{M}^{(0)}$ and $\mathcal{M}^{(1)}$ denoting the Born and one-loop
amplitudes. For each spin configuration in Sec.~\ref{sec:spin}, we
perform the spin sums using the appropriate fermion density matrices
and gauge-boson polarization projectors.\footnote{{In
the \software{FeynCalc} and \software{Tracer} routes, we contract the
physical gluon polarization tensors before expanding the Dirac traces.
This limits intermediate expression growth in the $qg\to q+X$
real-emission calculation.}}

Starting from the same generated amplitudes, we evaluate the spin and
color algebra separately with \software{FeynCalc}~\cite{Shtabovenko:2023idz},
\software{FORM}/\allowbreak\software{TFORM}~\cite{Ruijl:2017dtg,Davies:2026cci}, and
\software{Tracer}~\cite{Jamin:1991dp}. Traces involving $\gamma_5$ are
evaluated in the HVBM prescription~\cite{tHooft:1972tcz,Breitenlohner:1977hr}:
\software{FeynCalc} implements this prescription in its \texttt{BMHV} mode,
\software{FORM} constructs $\gamma_5$ from the
four-dimensional Levi-Civita tensor, and \software{Tracer} separates the physical
and evanescent subspaces. We express the three results in the same scalar
invariants with a common normalization. The trace coefficients retain
their dependence on $D=4-2\epsilon$, and powers of evanescent scalar products
are kept explicit for integration.

\subsection{Reverse unitarity and IBP reduction}
\label{sec:ibp}

For real emission at fixed tagged momentum, we integrate over the
unobserved massless pair with total momentum $P_X$ and invariant mass
$P_X^2=s_4$. Here $P_X=l+k-p$ for lepton--parton scattering and
$P_X=k_a+k_b-p$ for parton--parton scattering:
\begin{equation}
 \int\dd\Phi_2^{(D)}(P_X;r,P_X-r)\,F(r)
 =(2\pi)^{2-D}\int\dd^Dr\,
 \delta_+(r^2)\delta_+((P_X-r)^2)F(r),
 \label{eq:workflow-cut-measure}
\end{equation}
where $\delta_+(k^2)=\theta(k^0)\delta(k^2)$.
{Reverse unitarity was developed for inclusive
phase-space integrals~\cite{Anastasiou:2002yz} and extended to
differential distributions~\cite{Anastasiou:2003yy,Anastasiou:2003ds}.
It represents the on-shell constraints by cut propagators using the
Cutkosky relation~\cite{Cutkosky:1960sp},}
\begin{equation}
 \delta(X)=\frac{1}{2\pi i}
 \left(\frac{1}{X-i0}-\frac{1}{X+i0}\right).
 \label{eq:workflow-cut-definition}
\end{equation}
The positive-energy restrictions are retained in the physical
definition of the cut masters. The resulting cut integrals are reduced
by integration-by-parts (IBP)
identities~\cite{Chetyrkin:1981qh,Tkachov:1981wb} in the same way as loop
integrals.

We use partial fractions to rewrite the real-emission integrand as a sum
of terms with at most two linearly independent ordinary denominators.
The integral families contain the two cut propagators of
Eq.~\eqref{eq:workflow-cut-measure} and two ordinary denominators.
Integrals with a nonpositive power of either cut propagator vanish.
The spin projections change the scalar numerators while leaving these
denominator families unchanged.
\software{Kira}~\cite{Klappert:2020nbg,Lange:2025fba} generates the
reduction identities at a symbolic integration dimension $d$, using
the Laporta algorithm~\cite{Laporta:2000dsw}. The reduction coefficients
are rational functions of $d$ and the kinematic invariants. For each
family, we apply these identities to all spin projections, evaluating
them at $d=D$ for ordinary integrals and at the shifted dimensions
specified in Sec.~\ref{sec:dimshift} for integrals containing
evanescent numerators. The resulting master basis is common to the
different spin projections.

The real-emission masters are the two-body phase-space volume and
angular integrals containing both ordinary denominators. Up to
kinematic prefactors, the angular masters are values of a common
dimension-dependent function at different cross ratios. The virtual
integrals reduce to an off-shell massless bubble in the lepton--quark
process and to massless bubbles and boxes in the hadronic processes.

\Needspace{6\baselineskip}
\subsection{{Dimensional shifts} and analytic integration}
\label{sec:dimshift}

{We use {dimensional-shift} identities to express integrals
containing the evanescent numerator powers generated by the HVBM algebra
as scalar integrals in higher dimensions. Early one-loop formulations
include tensor reduction~\cite{Davydychev:1991va} and {dimensional-shift}
relations for scalar integrals~\cite{Bern:1992em,Bern:1993kr}. General
recurrence relations in the space-time dimension, including multiloop
integrals, were developed in
Refs.~\cite{Tarasov:1996br,Tarasov:1997kx}.} For a scalar
integrand $F(q)$ with four-dimensional external momenta, define
$\mu_q^2=-\widehat q^{\,2}$ in the mostly-minus metric. The normalized
{one-loop identity follows from
Ref.~\cite{Bern:1995db}:}
\begin{equation}
 \int\frac{\dd^Dq}{i\pi^{D/2}}(\mu_q^2)^n F(q)
 =(-\epsilon)_n
 \int\frac{\dd^{D+2n}q}{i\pi^{(D+2n)/2}}F(q),
 \qquad D=4-2\epsilon,
 \label{eq:workflow-dimension-shift}
\end{equation}
where $n$ is a nonnegative integer and
$(-\epsilon)_n=\Gamma(n-\epsilon)/\Gamma(-\epsilon)$.
For an insertion $(\widehat q^{\,2})^n$, the coefficient of the shifted
integral is $(-1)^n(-\epsilon)_n$, equal to $+\epsilon$ for $n=1$.
For the physical two-body phase-space measure in
Eq.~\eqref{eq:workflow-cut-measure}, the corresponding factor is
$(4\pi)^n(-1)^n(-\epsilon)_n$.

For the shifted integral, the IBP coefficients and master integrals are
evaluated at $d=D+2n$, while $D=4-2\epsilon$ is kept in the trace
coefficients, spin projectors and spin averages. The evanescent numerator
factors are retained because their products with poles can remain finite.

Each master is expressed as an Euler integral, with its normalization,
scale factors and branch prescription fixed by the original loop or
phase-space integral.
\software{SubTropica}~\cite{Giroux:2026tgd} computes the Laurent expansions
of the resulting linearly reducible Euler integrals. Virtual masters
are evaluated in a Euclidean region and continued with the Feynman
prescription to the physical region before taking the real part of
the interference. The same normalized master integrals and analytic
continuation are used with all three algebra implementations.

Singular terms in the integrated real contribution take the form
$(1-w)^{-1-\lambda\epsilon}B_\lambda(w,\epsilon)$, with $B_\lambda$
regular at $w=1$. {We expand $B_\lambda$ in $\epsilon$ and use
Eq.~\eqref{eq:endpoint-expansion} for the distributional expansion of
$(1-w)^{-1-\lambda\epsilon}$.} Each master is expanded to the order required for the finite
term of its product with the reduction coefficients, measure factors
and endpoint terms. Terms of positive order in $\epsilon$ can contribute
to the finite result when multiplied by poles in the remaining factors.

\subsection{Finite hard functions and physical checks}
\label{sec:gates}

Applying Eq.~\eqref{eq:endpoint-expansion} to the integrated real
contribution yields the coefficients of $\delta(1-w)$, $\cD_0$ and
$\cD_1$, together with the regular remainder. Smooth prefactors are
evaluated at $w=1$ in the distribution coefficients, with the remainders
from the plus-distribution terms included in the regular part according
to Eq.~\eqref{eq:distribution-form}. We then add the virtual contribution
and the $\msbar$ counterterms for ultraviolet renormalization and
initial- and final-state collinear subtraction. For lepton scattering,
the photon-in-lepton counterterm of Sec.~\ref{sec:observables} is added as
well. The lower-order
hard functions entering these counterterms are computed in $D$
dimensions; their $\epsilon$-dependent terms multiply collinear poles
and contribute to the finite coefficients.

For longitudinal polarization, the finite HVBM restoration term of
Eq.~\eqref{eq:hvbm-restoration} is added to the collinear subtraction.
For transversely polarized quarks, the collinear counterterms are
constructed with the transversity kernel $P^T_{qq}$. For each spin
configuration, summing the real, virtual and counterterm contributions
gives the finite hard function, with coefficients $A$, $B$, $C$ and $R$
defined in Eq.~\eqref{eq:distribution-form}.

{Gauge invariance is tested through Ward identities for the TMD currents
and by checking that selected hadronic real-emission squared matrix
elements are independent of the gluon polarization reference vectors,
as described in Appendix~\ref{app:acceptance}.}
The double and single poles cancel separately in the delta,
plus-distribution and regular sectors. We also check the
{RG evolution} of the finite hard functions. In the NLO
factorized cross section, {this evolution} cancels the terms generated by
the {RG evolution} of the incoming
parton distributions and {the final-state} fragmentation functions. We compute these
evolution terms by convolving the Born hard functions with the
leading-order splitting kernels, and verify the cancellation for each
factorization scale. In the
pure-annihilation channel $q_i\bar q_i\to q_k+X$, $k\ne i$, massless
helicity conservation gives
$\widehat\sigma_{LL}=-\widehat\sigma$ in each distribution sector,
providing a check of the longitudinal finite subtraction. Checks of
selected unresolved limits and master integrals are
described in Appendix~\ref{app:acceptance}. In the TMD calculation of
Sec.~\ref{sec:tmd}, the dependence on the auxiliary rapidity scale
cancels between the beam and soft contributions to the finite matching
coefficients.

Table~\ref{tab:validation} shows how the results obtained with the three
algebra programs compare with one another and with published results. The results
for transverse spin transfer in proton--proton collisions are also
checked through the crossing relations of
Sec.~\ref{sec:pp-transfer-crossing} and an independent implementation
of the finite assembly. All three algebra programs start from the same
generated amplitudes. Their results are processed with the same
scalar-reduction and assembly programs, master integrals and subtraction
conventions. The regular, plus-distribution
and delta coefficients are compared separately in the same
normalization.

\begin{table}[htbp]
 \centering
 \footnotesize
 \setlength{\tabcolsep}{4pt}
 \renewcommand{\arraystretch}{1.12}
 \begin{tabular}{@{}L{.19\linewidth}L{.25\linewidth}L{.13\linewidth}L{.37\linewidth}@{}}
 \toprule
 & \multicolumn{2}{c}{\software{FeynCalc} / \software{FORM} / \software{Tracer}} & \\
 \cmidrule(lr){2-3}
 Calculation & Quantities compared & Method & Other checks\\
 \midrule
 $pp$: UU and LL
 & Matrix elements before integration; finite NLO coefficients
 & Test points
 & Published UU/LL coefficients (test points); unpolarized
 $qg\to q+X$ endpoint coefficients (symbolic)\\[5pt]
 Direct $\ell p$: UU and LL
 & Finite NLO coefficients
 & Symbolic
 & Published coefficients (symbolic); {RG evolution checked against
 independently evaluated splitting-kernel convolutions}\\[5pt]
 $pp$: incoming TT
 & Matrix elements before integration; finite NLO coefficients$^{a}$
 & Test points
 & Published incoming-TT coefficients (test points)\\[5pt]
 $pp$: transfer TT
 & Matrix elements before integration; finite NLO coefficients$^{a}$
 & Test points
 & Crossing: symbolic at Born level, test points at NLO.
 Separate finite assembly (symbolic)$^{b}$\\[5pt]
 $\ell p$: transfer TT
 & Finite NLO coefficients
 & Symbolic
 & Born normalization, pole cancellation and {RG evolution}\\[5pt]
 TMD matching and soft functions
 & Beam and TMDPDF matching coefficients; soft functions
 & Symbolic
 & Published coefficients and soft functions (symbolic);
 cancellation of rapidity-scale dependence\\
 \bottomrule
 \end{tabular}
 \caption{Checks of the NLO calculations. The two central columns
 describe comparisons between results obtained with \software{FeynCalc}, \software{FORM} and
 \software{Tracer}. ``Symbolic'' denotes equality at general kinematics;
 ``test points'' denotes comparison at selected kinematic values.
 UU and LL denote unpolarized and double-longitudinal scattering.}
 \label{tab:validation}
 \begin{minipage}{\linewidth}
 \footnotesize
 \textsuperscript{a}\,{Finite-coefficient comparisons
 cover all five incoming-TT classes and all seven transfer-TT classes.
 For each class, the reduction and finite assembly are performed
 separately for each algebra route.}
 \par\smallskip
 \textsuperscript{b}\,The separate transfer assembly uses the same $D$-dimensional Born
 hard functions, reduced integrals and master values, with independently
 implemented endpoint expansion, coupling renormalization and collinear
 subtraction.
 \end{minipage}
\end{table}

\section{Reproduction of known NLO results}
\label{sec:benchmarks}
\label{sec:agreement}

With the \software{FeynFacet} workflow, we reproduce the NLO single-inclusive hard functions for unpolarized and
polarized $pp$ and $\ell p$ scattering, and the unpolarized quark and gluon
TMD soft functions and matching coefficients. We express the collinear
hard functions in the spin and normalization conventions of
Sec.~\ref{sec:physics} and the distribution basis of
Eq.~\eqref{eq:distribution-form}. Table~\ref{tab:validation} specifies
the symbolic and test-point comparisons.

\subsection{Unpolarized proton--proton scattering}
\label{sec:pp-uu}
\label{sec:coverage}

For the spin-averaged process $pp\to h+X$, we reproduce the
single-inclusive partonic hard functions {of
Refs.~\cite{Aversa:1988vb,Jager:2002xm}} through $\alpha_s^3$. The observed hadron is
unpolarized, and the parton that fragments into it is tagged while the
unobserved final states are {summed over}. Table~\ref{tab:full-pp-coverage}
lists the 16 tagged channel classes. The flavor sum in
Eq.~\eqref{eq:pp-factorization} also includes their charge-conjugate
channels and incoming-leg permutations.

\begin{table}[htbp]
\centering
\small
\begin{tabular}{ccc@{\qquad}ccc}
\toprule
Class & Tagged channel & First order & Class & Tagged channel & First order\\
\midrule
1 & $q_iq_k\to q_i+X$ & $\alpha_s^2$ & 9 & $q_i g\to\bar q_k+X$ & $\alpha_s^3$\\
2 & $q_iq_k\to g+X$ & $\alpha_s^3$ & 10 & $q_i g\to\bar q_i+X$ & $\alpha_s^3$\\
3 & $q_i\bar q_k\to q_i+X$ & $\alpha_s^2$ & 11 & $q_i\bar q_i\to q_i+X$ & $\alpha_s^2$\\
4 & $q_i\bar q_k\to g+X$ & $\alpha_s^3$ & 12 & $q_i\bar q_i\to g+X$ & $\alpha_s^2$\\
5 & $q_i\bar q_i\to q_k+X$ & $\alpha_s^2$ & 13 & $q_i g\to q_i+X$ & $\alpha_s^2$\\
6 & $q_iq_i\to q_i+X$ & $\alpha_s^2$ & 14 & $q_i g\to g+X$ & $\alpha_s^2$\\
7 & $q_iq_i\to g+X$ & $\alpha_s^3$ & 15 & $gg\to g+X$ & $\alpha_s^2$\\
8 & $q_i g\to q_k+X$ & $\alpha_s^3$ & 16 & $gg\to q_i+X$ & $\alpha_s^2$\\
\bottomrule
\end{tabular}
\caption{Tagged-parton channel classes for the UU and LL hadronic
calculations, with $k\ne i$. The incoming-TT channels are listed in
Sec.~\ref{sec:pp-tt}. Fixed flavor labels are not summed; sums over
unobserved flavors appear through $N_f$. These class numbers identify
the hard functions throughout the paper.}
\label{tab:full-pp-coverage}
\end{table}

Ten classes have an $\alpha_s^2$ Born contribution and receive virtual and
real-emission corrections, together with collinear counterterms, at
$\alpha_s^3$. Classes
2, 4, 7, 8, 9 and 10 first occur at $\alpha_s^3$, through real emission
and collinear subtraction. Their finite hard functions contain only the
regular term $R(v,w)$; the coefficients of $\delta(1-w)$, $\cD_0$ and
$\cD_1$ vanish. The analytic results retain $N_c$, $N_f$ and independent
renormalization, initial-state factorization and fragmentation scales.

After matching the tagged-parton and distribution conventions, the Born
terms and the four NLO coefficients agree at the comparison points with
{the unpolarized reference
implementation~\cite{Aversa:1988vb}} for all 16 classes. For
$qg\to q+X$, the Born term and the three endpoint coefficients also
agree symbolically with the published analytic expressions, after
translating their finite PDF and fragmentation-function subtraction terms
to our $\msbar$ convention. These identities hold for general $v$,
$N_c$, $N_f$ and independent scale logarithms.

\FloatBarrier
\subsection{Longitudinally polarized proton--proton scattering}
\label{sec:pp-ll}

For two longitudinally polarized protons, the hard functions
$\widehat\sigma_{LL}^{\,ab\to c}$ multiply
$g_1^{a/p}g_1^{b/p}D_1^{h/c}$ in the factorized cross section. We
reproduce the NLO coefficients {of
Ref.~\cite{Jager:2002xm}} for the same 16 classes in
Table~\ref{tab:full-pp-coverage}, with the incoming half-difference
normalization of Sec.~\ref{sec:spin} and an unpolarized observed hadron.
The separation into channels present at Born level and channels first
appearing at $\alpha_s^3$ is unchanged.

The initial-state collinear counterterms use helicity-dependent
splitting kernels on both incoming legs, whereas the fragmentation
counterterms use the unpolarized kernels. On each longitudinally
polarized incoming quark or antiquark leg, we include the finite HVBM
restoration term in Eq.~\eqref{eq:hvbm-restoration} to express the hard
functions in the polarized $\msbar$ convention of
Ref.~\cite{Jager:2002xm}. The Born terms and all four NLO distribution
coefficients agree at the comparison points, with the renormalization,
PDF and fragmentation scales chosen independently.

\subsection{Incoming TT in proton--proton scattering}
\label{sec:pp-tt}

For $p^\uparrow p^\uparrow\to h+X$ with an unpolarized observed
hadron, we reproduce the NLO hard functions {of
Ref.~\cite{Mukherjee:2005rw}}. The incoming-TT hard functions
defined in Sec.~\ref{sec:spin} multiply
$h_1^{a/p}h_1^{b/p}D_1^{h/c}$ in the factorized cross section. For a
common incoming transverse-spin axis, the spin-dependent partonic
cross section varies as $\cos(2\Phi)$, where $\Phi$ is the azimuth of
the tagged parton measured from that axis. Suppressing the other
kinematic arguments, we write
\begin{equation}
 \widehat\sigma_{TT}(\Phi)=\cos(2\Phi)\,H_{TT},
 \qquad
 H_{TT}=\frac1\pi\int_0^{2\pi}\dd\Phi\,
 \cos(2\Phi)\,\widehat\sigma_{TT}(\Phi).
 \label{eq:tt-harmonic-integral}
\end{equation}
The quantity compared with the reference result is $H_{TT}$, the
coefficient of $\cos(2\Phi)$. The angular weight in
Eq.~\eqref{eq:tt-harmonic-integral} extracts this coefficient; an
unweighted azimuthal integral vanishes.

We calculate $H_{TT}$ from two spin configurations. The $X$ axis lies
along the tagged parton's transverse momentum in the scattering plane,
and the $N$ axis is normal to that plane. We denote the finite hard functions
with both incoming spin axes along $X$ or along $N$ by
$\widehat\sigma^{\rm in}_{TT,XX}$ and
$\widehat\sigma^{\rm in}_{TT,NN}$, respectively. They use the invariant
normalization of Sec.~\ref{sec:observables} and the incoming-spin
difference in Table~\ref{tab:spin-conventions}. Resolving the common
spin axis along $X$ and $N$ gives the contributions
$\widehat\sigma^{\rm in}_{TT,XX}\cos^2\Phi$ and
$\widehat\sigma^{\rm in}_{TT,NN}\sin^2\Phi$;
the mixed components vanish by parity after integration over the
unobserved final state. Since
$\cos^2\Phi=(1+\cos(2\Phi))/2$ and
$\sin^2\Phi=(1-\cos(2\Phi))/2$, the coefficient of $\cos(2\Phi)$ is
\begin{equation}
 H_{TT}=\frac{\widehat\sigma^{\rm in}_{TT,XX}
                 -\widehat\sigma^{\rm in}_{TT,NN}}2.
 \label{eq:pp-tt-projections}
\end{equation}
Ref.~\cite{Mukherjee:2005rw} performs the angular projection using
the covariant expression
\begin{equation}
 {\cal F}(p,s_a,s_b)=\frac{s}{tu}
 \left[2(p\cdot s_a)(p\cdot s_b)
       +\frac{tu}{s}s_a\cdot s_b\right],
 \label{eq:pp-tt-projector}
\end{equation}
where $p$ is the tagged parton's momentum and $s_a,s_b$ are the
incoming spin four-vectors. Denote the tagged parton's transverse
momentum by $p_{T,c}$. For aligned unit transverse spins in the
partonic center-of-mass frame,
$s_a\cdot s_b=-1$ and
$p\cdot s_a=p\cdot s_b=-p_{T,c}\cos\Phi$.
Using $p_{T,c}^2=tu/s$, Eq.~\eqref{eq:pp-tt-projector} reduces to
\[
 {\cal F}=2\cos^2\Phi-1=\cos(2\Phi).
\]
The weight ${\cal F}/\pi$ therefore gives the angular projection in
Eq.~\eqref{eq:tt-harmonic-integral} using only Lorentz scalar products.
It extracts the same $H_{TT}$ as the spin-component difference in
Eq.~\eqref{eq:pp-tt-projections}, establishing the correspondence
between our calculation and the reference projection method.

The nonzero tagged channels are
\begin{equation}
 \begin{gathered}
 q_i\bar q_i\to q_k+X\quad(k\ne i),\\
 q_iq_i\to q_i+X,\\
 q_iq_i\to g+X,\\
 q_i\bar q_i\to q_i+X,\\
 q_i\bar q_i\to g+X.
 \end{gathered}
 \label{eq:pp-tt-channels}
\end{equation}
corresponding to classes 5, 6, 7, 11 and 12 in
Table~\ref{tab:full-pp-coverage}, together with the charge-conjugate
and incoming-leg assignments. The absence of leading-twist collinear
gluon transversity in a spin-$\tfrac12$ proton restricts the incoming
partons to quarks and antiquarks; the observed unpolarized parton can
be a gluon. The channel $q_iq_i\to g+X$ first contributes at
$\alpha_s^3$ and has no endpoint distributions.

The initial-state collinear counterterms use the transversity kernel
$P^T_{qq}$ on both incoming legs, while the final-state counterterms use
the unpolarized splitting kernels appropriate to $D_1$. After conversion
to the spin, azimuthal and invariant normalization used here, the Born
terms and the regular, $\cD_0$, $\cD_1$ and delta coefficients agree
at the comparison points with Ref.~\cite{Mukherjee:2005rw} for all five
channels. The corresponding incoming-TT amplitudes also enter the
crossing relations in Sec.~\ref{sec:pp-transfer-crossing}.

\subsection{Unpolarized lepton--proton scattering}
\label{sec:lp-uu}

For $\ell p\to h+X$ with an unobserved scattered lepton and an
unpolarized observed hadron, we reproduce the NLO hard functions
{of Ref.~\cite{Hinderer:2015hra}}. The hadron
transverse momentum sets the hard scale. The direct-lepton calculation
contains the three channels $q\to q$, $q\to g$ and $g\to q$,
including their charge conjugates. The diagonal channel receives the
Born contribution at $\alpha_{\rm em}^2$, its NLO virtual and real-emission
corrections, and the collinear counterterms. The off-diagonal channels begin at
$\alpha_{\rm em}^2\alpha_s$ and contain only the regular term. Their
QCD collinear subtractions involve final-state quark-to-gluon splitting
for $q\to g$ and incoming gluon-to-quark splitting for $g\to q$.

Integration over the scattered lepton includes the region in which the
exchanged photon becomes collinear to the incoming lepton. In the
massless-lepton calculation, this singularity is absorbed into
$f_{\gamma/\ell}$ in Eq.~\eqref{eq:lN-factorization}. The photon
counterterm uses $P_{\gamma/\ell}$ and the dimensionally regulated
Born hard functions for $\gamma q\to q+X$, $\gamma q\to g+X$ and
$\gamma g\to q+X$. Their $\epsilon$-dependent terms contribute to
the finite direct-lepton subtraction. Table~\ref{tab:ep-coverage}
lists the direct-lepton coefficients and the photon Born hard functions
computed alongside them. The supplemental resolved-photon hard
functions coincide with the Born coefficients of the ten
$\alpha_s^2$ QCD channels in Table~\ref{tab:full-pp-coverage}.

\begin{table}[htbp]
\centering
\small
\begin{tabular}{@{}C{0.20\linewidth}C{0.48\linewidth}C{0.23\linewidth}@{}}
\toprule
Contribution & Tagged channel basis & Hard-scattering accuracy\\
\midrule
Direct lepton & $\ell q\to q+X$, $\ell q\to g+X$, $\ell g\to q+X$
 & Through $\alpha_{\rm em}^2\alpha_s$\\[4pt]
Direct photon & $\gamma q\to q+X$, $\gamma q\to g+X$, $\gamma g\to q+X$
 & Born: $\alpha_{\rm em}\alpha_s$\\[4pt]
Resolved photon & $a_p b_\gamma\to c+X$: the ten nonzero QCD Born
classes in Table~\ref{tab:full-pp-coverage}
 & Born: $\alpha_s^2$\\
\bottomrule
\end{tabular}
\caption{Hard functions for single-inclusive lepton--proton scattering
with the scattered lepton unobserved. Both UU and LL are included, with
unpolarized final-state fragmentation and the corresponding antiquark
channels. The photon orders refer to the hard subprocesses before
convolution with the photon-in-lepton distribution and, for the resolved
term, the photon PDFs.}
\label{tab:ep-coverage}
\end{table}

For the direct-lepton comparisons we set
$\mu_R=\mu_F=\mu_F'=\mu_L=\mu$. After expressing the published
results in the endpoint-prefactor convention of
Eq.~\eqref{eq:distribution-form}, the four diagonal coefficients and
the two off-diagonal regular coefficients agree symbolically with
Ref.~\cite{Hinderer:2015hra}.

\FloatBarrier
\subsection{Longitudinally polarized lepton--proton scattering}
\label{sec:lp-ll}

With both the incoming lepton and proton longitudinally polarized, the
corresponding hard functions multiply $g_1^{i/p}D_1^{h/f}$ in the
polarized version of Eq.~\eqref{eq:lN-factorization}, with the helicity
difference specified in Sec.~\ref{sec:spin}. We reproduce the NLO results
{of Ref.~\cite{Hinderer:2017ntk}} for the same three
direct-lepton channels in Table~\ref{tab:ep-coverage}. The scattered
lepton remains unobserved, and the final-state fragmentation is
unpolarized.

The initial-state counterterms use the polarized quark and gluon
splitting kernels, while the fragmentation counterterms remain
unpolarized. The diagonal $q\to q$ channel includes the finite
incoming-quark restoration term of Eq.~\eqref{eq:hvbm-restoration}.
The photon-collinear subtraction uses the polarized photon-in-lepton
kernel $\Delta P_{\gamma/\ell}(y)=2-y$ and the corresponding polarized
photon Born hard functions.

With the common-scale and distribution conventions used in
Sec.~\ref{sec:lp-uu}, all four diagonal coefficients and both
off-diagonal regular coefficients agree symbolically with
Ref.~\cite{Hinderer:2017ntk}. For both UU and LL, the common-scale
dependence also agrees with that obtained from independently evaluated
incoming-parton, fragmentation and photon-in-lepton splitting-kernel
convolutions.

\subsection{Quark and gluon TMD soft functions}
\label{sec:tmd-soft}

The quark and gluon soft functions are obtained from the vacuum
Wilson-line matrix element in Eq.~\eqref{eq:tmd-soft-cut-operator},
with fundamental and adjoint color representations, respectively.
We evaluate this matrix element in the exponential-regulator convention
of Refs.~\cite{Li:2016ctv,Luo:2020epw}. With the tree-level result
normalized to unity and $a_s=\alpha_s(\mu)/(4\pi)$, we write
\begin{equation}
 S_i(\bT,\mu,\nu)
 =1+a_s S_i^{(1)}(\bT,\mu,\nu)+\order{a_s^2},
 \qquad C_q=C_F,\qquad C_g=C_A.
 \label{eq:tmd-soft-expansion}
\end{equation}

At one loop, interference between emissions from the two lightlike
Wilson lines gives the squared eikonal current $4g_s^2C_i/(k^+k^-)$.
The virtual integrals are scaleless and vanish in dimensional
regularization.
The real contribution is integrated with the transverse Fourier weight
and exponential regulator of Sec.~\ref{sec:tmd-setup}. We retain the
leading small-$\tau$ terms, including logarithms, at fixed $\epsilon$
before expanding in $\epsilon$. The Euler representation and reduction
of this integral are given in Appendix~\ref{app:tmd-integrals}. With
$\nu=1/\tau$, the bare coefficient is
\begin{equation}
 S_{i,\mathrm{bare}}^{(1)}
 =\frac{4C_i}{\epsilon^2}-\frac{4C_i\Lnu}{\epsilon}
 -C_i\left(2\Lp^2+4\Lp\Lnu+2\zeta_2\right)
 +\order{\epsilon},
 \label{eq:tmd-soft-poles}
\end{equation}
where $\Lnu=\ln(\nu^2/\mu^2)$ and $\zeta_2=\pi^2/6$.
Subtracting the poles in $\msbar$ gives
\begin{equation}
 S_i^{(1)}(\bT,\mu,\nu)
 =-C_i\left(2\Lp^2+4\Lp\Lnu+2\zeta_2\right).
 \label{eq:tmd-soft}
\end{equation}
The substitution $C_F\to C_A$ relates the quark and gluon results,
including the finite constant $-2C_i\zeta_2$. Both expressions agree
analytically with the one-loop soft functions of
Refs.~\cite{Li:2016ctv,Luo:2020epw} in the conventions of
Sec.~\ref{sec:tmd-setup}.

\subsection{TMDPDF matching coefficients}
\label{sec:tmd}
\label{sec:tmd-beam}

We reproduce the NLO matching coefficients of the unpolarized quark
and gluon TMDPDFs in Refs.~\cite{Luo:2019hmp,Luo:2020epw} by calculating
the partonic beam matrix elements and combining their finite matching
coefficients with the soft functions above. The four channels are
$ij=qq,qg,gq,gg$, where $i$ labels the TMD operator and $j$ the external
parton. The $qg$ coefficient is given per quark flavor. The beam and
TMDPDF coefficients are defined by Eqs.~\eqref{eq:tmd-ope}
and~\eqref{eq:tmd-physical}, respectively.

\subsubsection{Real-emission kernels and finite boundary terms}
\label{sec:tmd-kernels}

We include emission from the partons and Wilson lines, together with
the {two-gluon contact term} in the gluon field-strength operator.
Figure~\ref{fig:tmd-operator-cuts} shows representative cut diagrams;
the Wilson-line amplitudes are described in
Appendix~\ref{sec:tmd-model}.
\begin{figure}[t]
 \centering
\begingroup
\tikzset{
  ff-tmd/gluon/.style={draw=black,line width=.65pt,/tikzfeynman/gluon},
  ff-tmd/wilson/.style={draw=black,line width=.45pt,
    double=white,double distance=1.7pt},
  ff-tmd/cut/.style={draw=black,line width=.5pt,
    dash pattern=on 1.3pt off 1.7pt}}
\begin{minipage}[t]{.46\linewidth}
\centering
\begin{tikzpicture}[scale=.82,baseline=0pt]
  \path[use as bounding box] (-2.5,-2.15) rectangle (2.5,1.65);
  \draw[ff-tmd/cut] (0,-1.6)--(0,1.6);
  \draw[ff-tmd/gluon] (-1.2,-1.225)--(-1.2,-.175);
  \draw[ff-tmd/gluon] (-1.2,-.175)--(-1.2,1.225);
  \draw[ff-tmd/gluon] (1.2,1.225)--(1.2,-.175);
  \draw[ff-tmd/gluon] (1.2,-.175)--(1.2,-1.225);
  \draw[ff-tmd/gluon] (-1.2,-.175)--(1.2,-.175);
  \fill (-1.2,-.175) circle (1.5pt);
  \fill (1.2,-.175) circle (1.5pt);
  \node at (0,-1.95) {(a) $\mathcal I_{gg}$};
\end{tikzpicture}
\end{minipage}\hfill
\begin{minipage}[t]{.46\linewidth}
\centering
\begin{tikzpicture}[scale=.82,baseline=0pt]
  \path[use as bounding box] (-2.5,-2.15) rectangle (2.5,1.65);
  \coordinate (O) at (-.9,0);
  \coordinate (B) at (.9,0);
  \coordinate (LN) at (-1.9,1.5);
  \coordinate (LB) at (-1.9,-1.5);
  \coordinate (RN) at (1.9,1.5);
  \coordinate (RB) at (1.9,-1.5);
  \coordinate (E) at (-1.45,.825);
  \coordinate (F) at (1.45,-.825);
  \draw[ff-tmd/cut] (0,-1.6)--(0,1.6);
  \draw[ff-tmd/wilson] (O)--(LN);
  \draw[ff-tmd/wilson] (LB)--(O);
  \draw[ff-tmd/wilson] (RN)--(B);
  \draw[ff-tmd/wilson] (B)--(RB);
  \draw[ff-tmd/gluon] (E)--(F);
  \foreach \point in {O,B,E,F}{\fill (\point) circle (1.5pt);}
  \node[anchor=east] at (-2,1.5) {$n$};
  \node[anchor=east] at (-2,-1.5) {$\bar n$};
  \node[anchor=west] at (2,1.5) {$n$};
  \node[anchor=west] at (2,-1.5) {$\bar n$};
  \node at (0,-1.95) {(b) $S_q$};
\end{tikzpicture}
\end{minipage}
\endgroup
 \caption{Representative real-emission contributions to (a) the gluon
 beam matrix element used to determine $\cI_{gg}$ and (b) the
 fundamental soft function $S_q$. Curly lines denote gluons and double
 lines denote Wilson lines. The vertical dashed lines indicate the
 cuts; operator projectors and the remaining gauge links are implicit.}
 \label{fig:tmd-operator-cuts}
\end{figure}
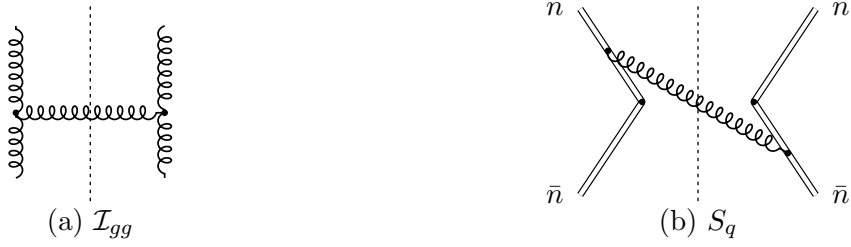
The partonic matrix elements yield the following unregularized
real-emission kernels for $0<\xi<1$, with $D=4-2\epsilon$ and the
$a_s$ normalization of Eq.~\eqref{eq:tmd-logs}:
\begin{align}
 P_{qq}^{\mathrm{real}}(\xi,\epsilon)
 &=2C_F\left[\frac{1+\xi^2}{1-\xi}
             -\epsilon(1-\xi)\right],\nonumber\\
 P_{qg}^{\mathrm{real}}(\xi,\epsilon)
 &=2T_F\left[1-\frac{2\xi(1-\xi)}{1-\epsilon}\right],\nonumber\\
 P_{gq}^{\mathrm{real}}(\xi,\epsilon)
 &=2C_F\left[\frac{1+(1-\xi)^2}{\xi}
             -\epsilon\xi\right],\nonumber\\
 P_{gg}^{\mathrm{real}}(\xi,\epsilon)
 &=4C_A\left[\frac{\xi}{1-\xi}
             +\frac{1-\xi}{\xi}+\xi(1-\xi)\right].
 \label{eq:tmd-real-kernels}
\end{align}
The factor $1/(1-\epsilon)$ in $P_{qg}^{\mathrm{real}}$ follows from
averaging over $D-2$ gluon polarizations. Away from $\xi=1$, the
transverse integral is $F_\perp=-1/\epsilon-\Lp+\order{\epsilon}$,
as derived in Appendix~\ref{app:tmd-integrals}. Its pole multiplies
the $\order{\epsilon}$ terms in Eq.~\eqref{eq:tmd-real-kernels} to
give the finite boundary terms
\begin{equation}
 I_{qq}^{(1)}(\xi)=2C_F(1-\xi),\qquad
 I_{qg}^{(1)}(\xi)=4T_F\xi(1-\xi),\qquad
 I_{gq}^{(1)}(\xi)=2C_F\xi,\qquad
 I_{gg}^{(1)}(\xi)=0.
 \label{eq:tmd-boundary}
\end{equation}
These functions equal the one-loop beam matching coefficients at
$\Lp=0$.

\subsubsection{\texorpdfstring{Soft-overlap}{Soft-overlap} subtraction and TMDPDF matching coefficients}
\label{sec:tmd-oracle}
\label{sec:tmd-status}
\label{sec:tmd-combination}

For on-shell external partons the virtual integrals are scaleless.
The overlap of the real-emission beam matrix element with the soft
region has support at $\xi=1$ in the diagonal channels. Division by
the bare soft function removes this overlap by subtracting
$\delta_{ij}\delta(1-\xi)S_{i,\mathrm{bare}}^{(1)}$ from the one-loop
coefficient of the unsubtracted beam matrix element. Matching onto
the renormalized collinear PDFs adds
$P_{ij}^{(0)}(\xi)/\epsilon$, where $P_{ij}^{(0)}$ are the full LO
splitting functions, including their endpoint distributions, in the
$a_s$ normalization. The remaining UV poles are proportional to
$\delta_{ij}\delta(1-\xi)$ and are removed by $\msbar$
renormalization of the beam operator. The finite beam coefficients are
\begin{align}
 \cI_{qq}^{(1)}(\xi)
 &=C_F\Bigl[2(1-\xi)+2\Lp(1+\xi)
       -4\Lp\cD_0(\xi)
       -2\Lp\LQ\delta(1-\xi)\Bigr],\nonumber\\
 \cI_{qg}^{(1)}(\xi)
 &=T_F\Bigl[4\xi(1-\xi)
       -2\Lp(1-2\xi+2\xi^2)\Bigr],\nonumber\\
 \cI_{gq}^{(1)}(\xi)
 &=C_F\left[2\xi
       -2\Lp\frac{2-2\xi+\xi^2}{\xi}\right],\nonumber\\
 \cI_{gg}^{(1)}(\xi)
 &=C_A\left[4\Lp\frac{-1+2\xi-\xi^2+\xi^3}{\xi}
       -4\Lp\cD_0(\xi)
       -2\Lp\LQ\delta(1-\xi)\right],
 \label{eq:tmd-nlo-kernels}
\end{align}
where $\cD_0(\xi)=[1/(1-\xi)]_+$ and the scale dependence is
displayed through $\Lp$ and $\LQ=2\ln(Q/\nu)$.

The TMDPDF matching coefficients follow by multiplying the beam
matching coefficients by the square root of the renormalized soft
function, as in
Eq.~\eqref{eq:tmd-physical}. At NLO this gives
\begin{equation}
 \cC_{ij}^{(1)}(\xi,\bT,\mu,\zeta)
 =\cI_{ij}^{(1)}(\xi,\bT,\mu,\nu;Q)
 +\frac12\delta_{ij}\delta(1-\xi)
 S_i^{(1)}(\bT,\mu,\nu).
 \label{eq:tmd-soft-combination}
\end{equation}
The beam and soft logarithms combine as
$\LQ+\Lnu=\ln(\zeta/\mu^2)$, so {the dependence on the auxiliary rapidity scale $\nu$}
cancels. Using Eq.~\eqref{eq:tmd-nlo-kernels}, the complete NLO TMDPDF
coefficients can be written as
\begin{equation}
 \cC_{ij}^{(1)}(\xi,\bT,\mu,\zeta)
 =\left.\cI_{ij}^{(1)}(\xi)\right|_{\LQ=\ln(\zeta/\mu^2)}
 -C_i\delta_{ij}\delta(1-\xi)
 \left(\Lp^2+\zeta_2\right).
 \label{eq:tmd-final-coefficients}
\end{equation}
The soft contribution changes only the diagonal endpoint terms at
this order; $\cC_{qg}^{(1)}=\cI_{qg}^{(1)}$ and
$\cC_{gq}^{(1)}=\cI_{gq}^{(1)}$. The beam coefficients in
Eq.~\eqref{eq:tmd-nlo-kernels} and the TMDPDF coefficients in
Eq.~\eqref{eq:tmd-final-coefficients} agree symbolically with the
corresponding one-loop expressions of Ref.~\cite{Luo:2020epw}.

\section{New NLO transverse-spin-transfer hard functions}
\label{sec:newresults}

We present the analytic NLO hard functions for transverse spin transfer
in proton--proton collisions (Sec.~\ref{sec:pp-transfer}) and in
lepton--proton collisions with the scattered lepton unobserved
(Sec.~\ref{sec:pheno}). The outgoing quark polarization can be probed
through polarized-hyperon production, Collins and dihadron fragmentation,
and spin-dependent energy correlators. These applications are discussed
in Sec.~\ref{sec:transfer-observables}.

\subsection{Proton--proton collisions}
\label{sec:pp-transfer}

The hadronic calculation determines the single-inclusive hard functions
for
\begin{equation}
 a^\uparrow(p_a)+b(p_b)\longrightarrow c^\uparrow(p)+X,
 \label{eq:pp-transfer-process}
\end{equation}
Here $p_a=k_a$ and $p_b=k_b$ in the notation of
Sec.~\ref{sec:observables}. The incoming parton $a$ and the observed
parton $c$ are quarks or
antiquarks, and $b$ is unpolarized. The hard function is differential
in the momentum $p$ of the observed parton $c$ and inclusive over the
unobserved final state $X$. The result covers seven tagged-parton classes through
$\alpha_s^3$: six have a Born contribution at $\alpha_s^2$, while
$q^\uparrow g\to\bar q^\uparrow+X$ first contributes at $\alpha_s^3$.
These results extend the Born spin-transfer coefficients used in
polarized fragmentation studies~\cite{Stratmann:1992gu,Collins:1993kq,Yuan:2007nd,Kang:2017btw}
to NLO.

\subsubsection{Spin projection and partonic channels}
\label{sec:pp-transfer-channels}

We use the projectors and full outgoing-spin difference of
Sec.~\ref{sec:spin}. Let $X_a$, $X_c$ and $N$ denote the physical unit
spin four-vectors along the transverse axes defined there. In the
partonic center-of-mass frame, $X_a$ points along the tagged parton's
transverse momentum, and the spatial part of $N$ points along
$\boldsymbol p_a\times\boldsymbol p$. With $s_a=s_T$ and $s_c=s_T'$,
we define
$\phi_a$ and $\phi_c$ by
$s_a=X_a\cos\phi_a+N\sin\phi_a$ and
$s_c=X_c\cos\phi_c+N\sin\phi_c$, so that each angle is measured from
its $X$ axis toward the common $N$ axis. The diagonal part of the
spin-projected squared matrix element is
\begin{equation}
 T_{\rm diag}(s_a,s_c)=T_{XX}\cos\phi_a\cos\phi_c
          +T_{NN}\sin\phi_a\sin\phi_c,
 \qquad T_{\rm C}=\frac{T_{XX}+T_{NN}}{2}.
 \label{eq:pp-transfer-tensor}
\end{equation}
The mixed $XN$ and $NX$ components are odd under reflection through
the scattering plane and vanish upon integration over the unobserved
momenta. We retain both diagonal components in $D=4-2\epsilon$
throughout the calculation.

The finite hard functions $\widehat\sigma_{T,XX}$ and
$\widehat\sigma_{T,NN}$ follow after phase-space integration,
renormalization and collinear subtraction in the normalization of
Eq.~\eqref{eq:pp-factorization}. Their Collins projection is
$\widehat\sigma_T=(\widehat\sigma_{T,XX}+\widehat\sigma_{T,NN})/2$.
We find $\widehat\sigma_{T,XX}=\widehat\sigma_{T,NN}$ in every channel.
The resulting spin-dependent partonic cross section is therefore
proportional to $\cos(\phi_a-\phi_c)$.

Table~\ref{tab:pp-transfer-coverage} lists the seven channels and the
incoming-TT processes used in the crossing comparison below. The first
incoming label identifies the parton carrying transversity. In
same-flavor annihilation, transfer to the quark and transfer to the
antiquark are distinct tagged processes. The table includes
both tagged processes, with $\bar q_i^\uparrow q_i\to q_i^\uparrow+X$
representing the charge conjugate of
$q_i^\uparrow\bar q_i\to\bar q_i^\uparrow+X$.

\begin{table}[htbp]
 \centering
 \small
 \begin{tabular}{cccc}
 \toprule
 Class & Transfer TT channel & First order & Incoming-TT class (tagged parton)\\
 \midrule
 1 & $q_i^\uparrow q_k\to q_i^\uparrow+X$ & $\alpha_s^2$ & $5\;(\bar q_k)$\\
 3 & $q_i^\uparrow\bar q_k\to q_i^\uparrow+X$ & $\alpha_s^2$ & $5\;(q_k)$\\
 6 & $q_i^\uparrow q_i\to q_i^\uparrow+X$ & $\alpha_s^2$ & $11\;(\bar q_i)$\\
 10 & $q_i^\uparrow g\to\bar q_i^\uparrow+X$ & $\alpha_s^3$ & $7\;(g)$\\
 11 & $q_i^\uparrow\bar q_i\to q_i^\uparrow+X$ & $\alpha_s^2$ & $11\;(q_i)$\\
 13 & $q_i^\uparrow g\to q_i^\uparrow+X$ & $\alpha_s^2$ & $12\;(g)$\\
 17 & $\bar q_i^\uparrow q_i\to q_i^\uparrow+X$ & $\alpha_s^2$ & $6\;(q_i)$\\
 \bottomrule
 \end{tabular}
 \caption{Transverse-spin-transfer channels, with $k\ne i$, calculated
 through $\alpha_s^3$ together with their charge conjugates. The last
 column gives the incoming-TT class of Eq.~\eqref{eq:pp-tt-channels}
 and the parton tagged before crossing. For classes 1 and 6, the
 antiquark-tagged incoming-TT channel is obtained from the quark-tagged
 representative by charge conjugation and interchange of its two
 incoming legs. For class 17, charge conjugation after crossing gives
 the displayed transfer channel.}
 \label{tab:pp-transfer-coverage}
\end{table}

The NLO-only contribution arises from the real process
$q_i^\uparrow g\to\bar q_i^\uparrow q_i q_i$.
The spin transfer is generated by interference between
amplitudes in which the two identical unobserved quarks are exchanged.
Quark-to-antiquark transfer also occurs in the Born annihilation
channel; the new real process adds a contribution with an unpolarized
incoming gluon.

\FloatBarrier
\subsubsection{Finite hard functions}
\label{sec:pp-transfer-hard}

We evaluate the real-emission contribution and the Born--one-loop
interference with the methods of Sec.~\ref{sec:workflow}. The soft poles
cancel between real emission and the renormalized Born--one-loop
interference. The remaining collinear poles are subtracted in the
$\msbar$ scheme with the transversity kernel for the polarized incoming
and observed fermions, and unpolarized splitting kernels for the
incoming parton $b$.
The counterterms retain the $D$-dimensional Born terms, whose
evanescent parts contribute to the finite result when multiplied by
collinear poles. The sum of real, virtual, renormalization and
collinear contributions is finite separately for $XX$ and $NN$.

{Let $c_T^{(0)}(v)$ denote the dimensionless Born coefficient for
the channel $ab\to c+X$.} In the normalization of
Eq.~\eqref{eq:pp-factorization}, the hard function for the Collins
projection is
\begin{align}
 \widehat\sigma_T^{\,ab\to c}
 ={}&\frac{\alpha_s^2}{s}\,{c_T^{(0)}(v)}\delta(1-w)\nonumber\\
 &+\frac{\alpha_s^3}{\pi s v}
 \Big[\frac{R_T(v,w)}{w}
       +B_T(v)\cD_0(w)+C_T(v)\cD_1(w)
       \nonumber\\
 &\hspace{25mm}+A_T(v)\delta(1-w)\Big]
 +\order{\alpha_s^4}.
 \label{eq:pp-transfer-finite}
\end{align}
Here $v$ and $w$ are defined in Eq.~\eqref{eq:vw}, and the channel
labels on the right-hand side are suppressed. The endpoint
distributions multiply the prefactor evaluated at $w=1$, whereas the
regular term retains $1/w$. We use $\alpha_s=\alpha_s(\mu_R)$,
a common scale $\mu_F$ for the two incoming PDFs, and an independent
fragmentation scale $\mu_F'$, with logarithms $L_R$, $L_F$ and $L_F'$
as defined in Table~\ref{tab:scales}.

For $q_i^\uparrow q_k\to q_i^\uparrow+X$ with $k\ne i$, the Born
coefficient is
\begin{equation}
 {c_T^{(0)}(v)}=\frac{N_c^2-1}{N_c^2(1-v)^2}.
 \label{eq:pp-transfer-example-born}
\end{equation}
{At Born kinematics, $t=-s(1-v)$ and $u=-sv$. The coefficient in
Eq.~\eqref{eq:pp-transfer-example-born} is related to the Collins hard
factor of Ref.~\cite{Yuan:2007nd} by}
\begin{equation}
 \frac{s^2}{\pi\alpha_s^2}
 {H_{qq'\to qq'}^{\rm Collins}}
 =-\frac{N_c^2-1}{N_c^2}\frac{su}{t^2}
 ={v\,c_T^{(0)}(v)}.
 \label{eq:pp-born-normalization-bridge}
\end{equation}
{The dimensionless hard factor on the left is also used in
Ref.~\cite{Kang:2017btw}. The additional factor $1/v$ in
$c_T^{(0)}$ arises from the Born phase-space constraint,}
$\delta(s+t+u)=\delta(1-w)/(sv)$.

Writing $\ell_v=\ln v$ and $\ell_{\bar v}=\ln(1-v)$, the two
plus-distribution coefficients for this channel are
\begin{align}
 C_T(v)&=5C_F\,{v\,c_T^{(0)}(v)},\nonumber\\
 B_T(v)&={v\,c_T^{(0)}(v)}\left[
 -C_F\left(\frac34+2L_F+L_F'\right)
 -2(C_A-C_F)\ell_{\bar v}
 +(11C_F-2C_A)\ell_v\right].
 \label{eq:pp-transfer-example-endpoints}
\end{align}
The complete delta-function coefficient is
\begin{align}
 \frac{A_T(v)}{{v\,c_T^{(0)}(v)}}
 ={}&\frac{11N_c-2N_f}{6}\,L_R
 +C_F\left[
 \left(\ell_{\bar v}-\ell_v-\frac32\right)L_F
 -\left(\ell_v+\frac34\right)L_F'\right]\nonumber\\
 &+\frac{225+115N_c^2-40N_cN_f
 +\pi^2\left[24+21N_c^2+9(N_c^2-2)v^2\right]}{72N_c}
 \nonumber\\
 &+\frac{6/v-18-7N_c^2+4N_cN_f+3(N_c^2-2)v}{12N_c}\,
 \ell_{\bar v}
 -\frac{1+N_c^2+2(N_c^2-2)v}{8N_c}\,\ell_v\nonumber\\
 &+\frac{2/v^2+2+5N_c^2+(N_c^2-2)v^2}{8N_c}\,
 \ell_{\bar v}^{\,2}
 +\frac{-34+19N_c^2+(N_c^2-2)v^2}{8N_c}\,\ell_v^2
 \nonumber\\
 &-\frac{9N_c^2-12+(N_c^2-2)v^2}{4N_c}\,
 \ell_v\ell_{\bar v}.
 \label{eq:pp-transfer-example-delta}
\end{align}
The regular coefficient for this channel and the complete results for
all seven classes are supplied in the ancillary file
\texttt{ProtonProtonTransfer.wl}, with the normalization and scale
conventions specified in Appendix~\ref{app:coefficients}.

The distinct-flavor quark and antiquark channels have the same Born
coefficient but different regular NLO terms. For
$q_i^\uparrow g\to\bar q_i^\uparrow+X$, the result instead has
${c_T^{(0)}}=A_T=B_T=C_T=0$ and consists entirely of the finite regular
coefficient. These contributions enter the flavor sums for the
fragmentation observables of Sec.~\ref{sec:transfer-observables}.

\subsubsection{Crossing from incoming transverse polarization}
\label{sec:pp-transfer-crossing}

The regulated transfer amplitudes and scalar integrands are related by
crossing to those of the incoming-TT calculation in
Sec.~\ref{sec:pp-tt}.
Crossing the polarized incoming leg $b$ and the spin-summed observed
leg $c$ relates
\begin{equation}
 a^\uparrow b^\uparrow\to c+X
 \quad\longleftrightarrow\quad
 a^\uparrow\bar c\to\bar b^\uparrow+X.
 \label{eq:pp-transfer-crossing-process}
\end{equation}
For the basic map $p_b\to-p$, $p\to-p_b$, the invariants obey
\begin{equation}
 s\leftrightarrow t,\qquad u\to u,\qquad
 v\to-\frac{v}{1-v},\qquad w\to w.
 \label{eq:pp-transfer-crossing-map}
\end{equation}
The required tagged incoming-TT channels and charge-conjugation operations are
specified in Table~\ref{tab:pp-transfer-coverage}. For classes 1 and 6,
{combining the interchange of the incoming TT legs with the basic
crossing gives} $(s,t,u)\to(t,u,s)$ on the canonical quark-tagged
incoming-TT expression.

Let $\mathcal C$ denote the crossing and species relabeling of the
unintegrated scalar expressions. With $T_{XX},T_{NN}$ as in
Eq.~\eqref{eq:pp-transfer-tensor} and $T_{XX}^{\rm in},T_{NN}^{\rm in}$
the corresponding components of the incoming TT squared matrix element,
the spin-axis transformations and the color and spin averages give
\begin{align}
 T_{XX}&=-\varsigma_{\bar c}r_{\bar c}(D)\,
              {\cal C}[T_{XX}^{\rm in}],\qquad
 T_{NN}=+\varsigma_{\bar c}r_{\bar c}(D)\,
              {\cal C}[T_{NN}^{\rm in}],\nonumber\\
 (\varsigma_{\bar c},r_{\bar c})&=
 \begin{cases}
 (+1,1),&\bar c=q,\bar q,\\
 (-1,[(D-2)C_F]^{-1}),&\bar c=g.
 \end{cases}
 \label{eq:pp-transfer-crossing-factors}
\end{align}
Here $\bar c$ is the unpolarized incoming parton in the transfer
process. For a gluon, $r_{\bar c}$ includes its color and
$D$-dimensional polarization average. We retain the full $D$ dependence
of the crossing factors until the Laurent expansion. The component relations map the
incoming-TT
projection $(T_{XX}^{\rm in}-T_{NN}^{\rm in})/2$ into the transfer
projection $(T_{XX}+T_{NN})/2$.

The crossed integrands are integrated in the physical transfer region
and combined with the transfer collinear counterterms. The loop
integrals are analytically continued to this region before taking the
real part of the Born--one-loop interference. The crossed and direct
calculations agree for all seven channels in the comparisons summarized
in Table~\ref{tab:validation} and Appendix~\ref{app:spec}.
An independent implementation of the endpoint expansion,
renormalization and collinear subtraction reproduces the regular and
endpoint distribution coefficients and the equality
$\widehat\sigma_{T,XX}=\widehat\sigma_{T,NN}$.

\subsection{Lepton--proton collisions}
\label{sec:pheno}

We calculate the analytic NLO hard function for
$\ell(l)q^\uparrow(k)\to q^\uparrow(p)+X$ through
$\alpha_{\rm em}^2\alpha_s$. The incoming lepton is unpolarized, and we
integrate over the momenta of the scattered lepton and the other
unobserved particles at fixed observed-quark momentum $p$. The large
transverse momentum of the observed quark sets the hard scale.

At the EIC, Collins hadron-in-jet measurements can probe quark transversity
distributions in the proton through the correlation between the proton's
transverse spin and the hadron's azimuthal angle around the jet axis.
Leading-order studies of the single-inclusive
process include quasireal-photon
contributions~\cite{DAlesio:2026pca,Flore:2026gtk}; the coefficient below
supplies the NLO QCD correction to the direct-lepton production
mechanism. These measurements complement the electron--jet observables
with a tagged scattered lepton and small electron--jet momentum imbalance proposed in
Ref.~\cite{Arratia:2020nxw}. Polarized-hyperon production provides another
application of transverse spin transfer, for which one-loop
leptoproduction corrections were studied in
Ref.~\cite{Contogouris:1996transverse}. The fragmentation observables are
developed in Sec.~\ref{sec:transfer-observables}.

\subsubsection{Spin projection and finite hard function}
\label{sec:new}

We use $s=2l\cdot k$, $t=-2k\cdot p$, $u=-2l\cdot p$, and
$v=1+t/s$, $w=-u/(s+t)$. The spin projectors, physical axes and
dimensional prescription are those of Sec.~\ref{sec:spin}. The
Collins projection of the spin-projected squared matrix element is
$T_{\rm C}=(T_{XX}+T_{NN})/2$.

At Born level,
\begin{equation}
 \frac{T_{NN}^{(0)}}{e^4e_q^2}=-\frac{4su}{t^2}-(D-4),\qquad
 \frac{T_{XX}^{(0)}}{e^4e_q^2}=-\frac{4su}{t^2}+(D-4).
 \label{eq:transverse-born-tensor}
\end{equation}
The evanescent terms cancel in $T_{\rm C}^{(0)}$, and
$T_{\rm C}^{(0)}/U^{(0)}=2v/(1+v^2)$ in four dimensions, where
$U^{(0)}$ is the spin-averaged Born squared matrix element.
The real-emission diagonal components also agree in four dimensions;
the evanescent contributions to their average are retained through
integration. Collinear subtraction uses the real kernel in
Eq.~\eqref{eq:transverse-D-kernel} and the $\msbar$ transversity
counterterms specified in Sec.~\ref{sec:spin}.

Using the methods of Sec.~\ref{sec:workflow}, we combine the real-emission
contribution from $\ell q^\uparrow\to\ell q^\uparrow g$ with the Born--one-loop
interference and the quark and lepton--photon collinear counterterms.
We compute the finite Collins combination
$\widehat\sigma_T^{\ell q}
=(\widehat\sigma_{T,XX}^{\ell q}+\widehat\sigma_{T,NN}^{\ell q})/2$.
In the single-inclusive normalization of
Eq.~\eqref{eq:lN-factorization}~\cite{Hinderer:2015hra}, it is
\begin{align}
 \widehat\sigma_T^{\ell q}
 ={}&\frac{4\alpha_{\rm em}^2e_q^2}{s(1-v)^2}
       \delta(1-w)\nonumber\\
 &+\frac{\alpha_{\rm em}^2e_q^2 C_F\alpha_s(\mu_R)}{\pi s v}
 \Big[\frac{R_T(v,w)}{w}
       +B_T(v)\cD_0(w)+C_T(v)\cD_1(w)\nonumber\\
 &\hspace{25mm}+A_T(v)\delta(1-w)\Big]
 +\order{\alpha_{\rm em}^2\alpha_s^2}.
 \label{eq:transverse-finite-hard}
\end{align}
The endpoint distributions multiply the prefactor evaluated at $w=1$;
the regular term retains $1/w$. With the scale logarithms of
Table~\ref{tab:scales}, the endpoint coefficients are
\begin{align}
 C_T(v)&=\frac{16v}{(1-v)^2},\nonumber\\
 B_T(v)&=\frac{4v}{(1-v)^2}
 \left[2\ln v+2\ln(1-v)-L_F-L_F'\right],\nonumber\\
 A_T(v)&=\frac{v}{(1-v)^2}
 \Big[-16+6\ln(1-v)-3L_F-3L_F'\nonumber\\
 &\hspace{22mm}+4\ln v\ln(1-v)+2\ln^2v-4L_F'\ln v\Big].
 \label{eq:transverse-endpoints}
\end{align}
The complete regular coefficient $R_T(v,w)$ is given in
Appendix~\ref{app:transverse-coefficients}, and all four coefficients are
supplied in \texttt{Collins.wl} in this endpoint-prefactor convention.
The regular term also contains $L_L=\ln(\mu_L^2/s)$ from the
photon-in-lepton subtraction.

\subsubsection{Photon-collinear subtraction and accompanying Born channels}

Integration over the scattered lepton includes the region in which it
is collinear to the incoming lepton and a nearly on-shell photon
initiates the hard scattering. We factorize this singularity into the
unpolarized $\msbar$ photon
distribution~\cite{Hinderer:2015hra},
\begin{equation}
 f_{\gamma/\ell}(y,\mu_L)
 =\frac{\alpha_{\rm em}}{2\pi}
  \frac{1+(1-y)^2}{y}
  \left[\ln\frac{\mu_L^2}{y^2m_\ell^2}-1\right]
  +\order{\alpha_{\rm em}^2},
 \label{eq:collins-photon-flux}
\end{equation}
where $y$ is the photon momentum fraction and $m_\ell$ is the lepton
mass. {The RG evolution of $f_{\gamma/\ell}$ over $\mu_L$ is given by}
$\partial f_{\gamma/\ell}/\partial\ln\mu_L^2
=\alpha_{\rm em}P_{\gamma/\ell}(y)/(2\pi)$, with
$P_{\gamma/\ell}(y)=[1+(1-y)^2]/y$.

The corresponding Born process is
$\gamma q^\uparrow\to q^\uparrow g$. Averaging over the photon
polarizations and the incoming-quark color gives the spin-projected
squared matrix element
\begin{equation}
 \frac{T_{\rm C}^{\gamma q}}{C_Fe^2e_q^2g_s^2}
 =4-\frac{2\epsilon^2t_\gamma^2}
          {(1-\epsilon)s_\gamma u_\gamma},
 \qquad
 (s_\gamma,t_\gamma,u_\gamma)=(ys,t,yu).
 \label{eq:transverse-photon-born}
\end{equation}
The Born constraint $ys+t+yu=0$ fixes the photon fraction. In terms of
this fraction, the polynomial $p_L$ multiplying $L_L$ in
Appendix~\ref{app:transverse-coefficients} satisfies
\begin{equation}
 y_\gamma=\frac{1-v}{1-vw},\qquad
 p_L=(1-vw)^2+v^2(1-w)^2
     =(1-vw)^2\big[1+(1-y_\gamma)^2\big],
 \label{eq:lepton-photon-polynomial-check}
\end{equation}
Since $T_{\rm C}^{\gamma q}/(C_Fe^2e_q^2g_s^2)=4$ in four
dimensions, the Born delta function reduces the photon convolution to
\begin{equation}
 \int_0^1\frac{\dd y}{y}\,
 f_{\gamma/\ell}(y,\mu_L)
 \widehat\sigma_T^{\,\gamma q,(0)}(ys,t,yu)
 =\frac{4\alpha_{\rm em}e_q^2C_F\alpha_s}{s(1-v)}
  f_{\gamma/\ell}(y_\gamma,\mu_L).
\end{equation}
{The $L_L$ term of the direct-lepton hard function determines its
RG evolution over $\mu_L$:}
\begin{align}
 \frac{\partial\widehat\sigma_T^{\ell q}}
      {\partial\ln\mu_L^2}
 &=-\frac{2\alpha_{\rm em}^2e_q^2C_F\alpha_s}{\pi s}
   \frac{p_L}{(1-v)^2(1-vw)}\nonumber\\
 &=-\frac{\partial}{\partial\ln\mu_L^2}
   \left[f_{\gamma/\ell}\otimes
         \widehat\sigma_T^{\,\gamma q,(0)}\right].
 \label{eq:lepton-photon-scale-check}
\end{align}
Thus the direct-lepton hard function and the photon distribution
convolved with the photon Born hard function have canceling $\mu_L$
dependence through $\alpha_{\rm em}^2\alpha_s$.

The double poles cancel between real emission and the Born--one-loop
interference, and the quark and photon counterterms cancel the remaining
single poles in each distribution coefficient. {We check the RG
evolution over each of the three factorization scales.}
{The RG evolution of the hard function over $\mu_F$ and $\mu_F'$
cancels the contributions from the evolution of the incoming-quark PDF
and the quark FF in the factorized cross section. These contributions
are calculated using $P^T_{qq}$. The RG evolution over $\mu_L$ cancels between
the direct-lepton and photon contributions, as shown in}
Eq.~\eqref{eq:lepton-photon-scale-check}.
The correction has no explicit $L_R$ term because
the direct-lepton Born process contains no strong coupling. All three
algebra implementations give identical finite direct-lepton hard
functions and the same photon Born tensor, as summarized in
Table~\ref{tab:validation}.

In the resolved-photon term, the QCD Born hard functions are convolved
with the unpolarized photon PDFs $f_{b/\gamma}(r,M_\gamma)$. For the
spin average and Collins projection, labelled $A=U,T$, respectively,
we define the production kernel
\begin{equation}
 {\cal K}_{ac}^{A}
 =\widehat\sigma_A^{\,\ell a\to c}
 +f_{\gamma/\ell}\otimes\widehat\sigma_A^{\,\gamma a\to c,(0)}
 +\sum_b f_{\gamma/\ell}\otimes f_{b/\gamma}
                    \otimes\widehat\sigma_A^{\,ab\to c,(0)}.
 \label{eq:collins-channel-operator}
\end{equation}
Here $\widehat\sigma_U\equiv\widehat\sigma$, $a$ is the proton parton,
$b$ is the resolved-photon parton, and
$c$ is tagged. The direct-lepton hard functions are retained through
NLO, while the direct- and resolved-photon hard functions enter at Born
level. The photon convolution integrates $\dd y/y$ and evaluates the
hard function at $(ys,t,yu)$; the resolved term additionally integrates
$\dd r/r$ and uses $(rys,t,ryu)$. In both cases $t=(k-p)^2$ is the
invariant between the proton parton and the tagged parton. The structure
functions in Eq.~\eqref{eq:collins-factorization} involve further
convolutions with the proton distributions and fragmenting jet
functions.

For the resolved contribution, write the Born spin-projected squared
matrix elements as
$\overline{|\mathcal M|^2}_{T}=g_s^4\mathcal T_{ab\to c}$, with the
proton parton $a$ listed first and the full outgoing-spin difference
projected onto $T_{\rm C}$. Here $s,t,u$ are the invariants of the
resolved Born subprocess, with $t$ formed from the proton parton and
the tagged parton. Defining $K=(N_c^2-1)/N_c^2$, the nonzero
four-dimensional representatives are
\begin{equation}
 \begin{array}{c|l}
  ab\to c & \mathcal T_{ab\to c}\\ \hline
  qq'\to q,\ q\bar q'\to q\quad(q'\ne q)
    &-Ksu/t^2\\
  qq\to q&-Ksu/t^2+Ks/(N_ct)\\
  q\bar q\to q&-Ksu/t^2+Ku/(N_ct)\\
  q\bar q\to\bar q&-K/N_c\\
  qg\to q&-2su/t^2+K
 \end{array}
 \label{eq:resolved-transverse-born}
\end{equation}
The charge-conjugate channels have the same reduced squared matrix elements, and transfer
to a different-flavor tagged fermion vanishes. Both physical diagonal
components give the entries in Eq.~\eqref{eq:resolved-transverse-born}.
The tagged identical-quark channel carries no additional symmetry
factor $1/2$.

\subsection{Applications for fragmentation and jet substructure}
\label{sec:transfer-observables}
\label{sec:ep-transfer-analyzers}

Transverse spin transfer can be probed through the polarization of an
identified hadron or through the angular distribution of unpolarized
hadrons. The corresponding {transverse spin analyzer} ${\cal A}^c$
enters the structure functions in proton--proton and lepton--proton
collisions as
\begin{align}
 F_T^{pp}[{\cal A}]
 &=\frac{1}{S}\sum_{a,c=q,\bar q}\sum_b
 h_1^{a/p}\otimes f_1^{b/p}\otimes
 \widehat\sigma_T^{\,ab\to c}\otimes{\cal A}^c,
 \label{eq:pp-transfer-analyzer}\\
 F_T^{\ell p}[{\cal A}]
 &=\frac{1}{S}\sum_{a,c=q,\bar q}
 h_1^{a/p}\otimes{\cal K}_{ac}^{T}\otimes{\cal A}^c.
 \label{eq:ep-transfer-analyzer}
\end{align}
These structure functions use the finite production hard functions
$\widehat\sigma_T$ derived from the Collins projection in
Secs.~\ref{sec:pp-transfer} and~\ref{sec:pheno}. The
convolution measures are those of Eqs.~\eqref{eq:pp-factorization}
and~\eqref{eq:lN-factorization}: $\dd x_a/x_a\,\dd x_b/x_b$ for the
incoming proton partons in $pp$, $\dd x/x$ for the proton parton in
$\ell p$, and $\dd z/z^2$ for the final state. The production kernel
${\cal K}_{ac}^{T}$ is defined in
Eq.~\eqref{eq:collins-channel-operator}; its photon-fraction
convolutions are already included. {The lepton--proton cross section is inclusive in the scattered lepton.}

\subsubsection{Jet conventions and perturbative matching}
\label{sec:jet-conventions}
\label{sec:jet-matching}

We use anti-$k_T$ jets of radius $R$ with four-momentum recombination,
so the standard jet axis points along the spatial momentum of the
recombined jet. Let $p_T$ and $\eta_J$ denote the measured jet transverse
momentum and rapidity. The jet carries the light-cone fraction
$z=p_J^-/p_c^-$ of its parent parton, whereas an identified hadron
carries the fraction $z_h=p_h^-/p_J^-$ of the jet; the large light-cone
components are defined along the jet direction. The final-state
convolution integrates over $z$, with $p_{T,c}=p_T/z$, while $z_h$ is
a measured fragmentation variable. The vector $\boldsymbol j_\perp$
is the hadron momentum transverse to the standard jet axis.

At leading power for $R\ll1$ and $p_T R\gg\Lambda_{\rm QCD}$, the
jet cross section factorizes into incoming PDFs, an inclusive partonic
hard function and a jet function~\cite{Kang:2016mcy,Lee:2024icn}.
The hard functions of Secs.~\ref{sec:pp-transfer} and~\ref{sec:pheno}
describe production of the energetic parent parton at the scale
$p_T$, while the jet function describes radiation at the lower scale
$p_T R$.
For transverse-momentum-dependent fragmentation, the hierarchy is
$j_\perp\ll p_T R\ll p_T$. The fragmenting jet function then also
describes the measured transverse momentum and soft recoil inside the
jet~\cite{Kang:2017btw,Kang:2023elg}. The incoming PDFs are collinear
because the event recoil is integrated over.

{For an identified hadron inside the jet,
${\cal J}_c$ is replaced by a fragmenting jet function in the
measured-jet convention of Eq.~\eqref{eq:jet-parent-convention}.}
We denote its unpolarized and Collins TMD components by
${\cal D}_1^{h/c}$ and ${\cal H}_1^{\perp,h/c}$, respectively. They
depend on $(z,z_h,j_\perp^2,p_T R,\mu,\zeta_J)$.
Soft radiation shifts the standard jet axis. {The measured transverse
momentum obeys
$\boldsymbol j_\perp=\boldsymbol k_\perp+z_h\boldsymbol\lambda_\perp$,
where $\boldsymbol k_\perp$ is the hadron momentum transverse to the
fragmenting parton and $\boldsymbol\lambda_\perp$ is the in-jet soft
recoil. In transverse coordinate space, the unsubtracted TMD
fragmentation function multiplies the in-jet soft function. At NLO
this product is the subtracted TMD fragmentation function at the
Collins--Soper scale $\zeta_J=(p_T R)^2$~\cite{Kang:2023elg}.}

\Needspace{10\baselineskip}
Radiation at the jet boundary is described by the matching coefficient
that relates the Collins fragmenting jet function to TMD fragmentation.
In the {measured-jet} convention, it is~\cite{Kang:2023elg}
\begin{align}
 \widehat H^T_{q\to q}(z,p_T R,\mu)
 ={}&\delta(1-z)+\frac{\alpha_s(\mu)}{2\pi}\bigg\{
 L_J P^T_{qq}(z)
 -4C_Fz\left[\frac{\ln(1-z)}{1-z}\right]_+\nonumber\\
 &\quad+C_F\left(-\frac{L_J^2}{2}-\frac32L_J
                  +\frac{\pi^2}{12}\right)\delta(1-z)\bigg\},
 \label{eq:collins-jet-matching}
\end{align}
where $P^T_{qq}(z)=C_F[2z/(1-z)_++\tfrac32\delta(1-z)]$.
With the arguments of ${\cal H}_1^{\perp,h/q}$ specified above,
the matching relation at this order is~\cite{Kang:2023elg}
\begin{equation}
 {\cal H}_1^{\perp,h/q}
 =\widehat H^T_{q\to q}(z,p_T R,\mu)\,
  H_1^{\perp,h/q}(z_h,j_\perp^2;\mu,\zeta_J),
 \qquad \zeta_J=(p_T R)^2,
 \label{eq:collins-fjf-matching}
\end{equation}
{where $H_1^{\perp,h/q}$ is the subtracted Collins TMD fragmentation
function evaluated at the measured transverse momentum
$\boldsymbol j_\perp$.}
The transverse coefficient above and the unpolarized quark and gluon
jet matching coefficients are established ingredients from
Ref.~\cite{Kang:2023elg}.

The production and jet corrections combine in a single perturbative
expansion. Suppressing the incoming distributions and flavor sums,
\begin{equation}
 [\widehat\sigma_T\otimes{\cal A}]_{\rm NLO}
 =\widehat\sigma_T^{(0)}\otimes{\cal A}^{(0)}
 +\widehat\sigma_T^{(1)}\otimes{\cal A}^{(0)}
 +\widehat\sigma_T^{(0)}\otimes{\cal A}^{(1)}.
 \label{eq:jet-nlo-assembly}
\end{equation}
The superscripts on ${\cal A}$ denote the leading and first
perturbative matching contributions, with the nonperturbative
fragmentation function retained in each term. The contributions
labelled $(1)$ include one additional power of $\alpha_s$. This gives the $pp$
contribution through $\alpha_s^3$ and the direct-lepton contribution
through $\alpha_{\rm em}^2\alpha_s$. At the photon accuracy specified
in Sec.~\ref{sec:pheno}, the direct- and resolved-photon Born hard
functions multiply the leading jet matching contribution.

\subsubsection{Hadron spin and azimuthal correlations}

The polarized proton defines the $+z$ direction, and the observed
hadron or jet lies in the $xz$ plane. The incoming and outgoing
transverse bases use the in-plane axis $X$ and the normal $N$ of
Sec.~\ref{sec:pp-transfer}; the outgoing $X$ axis points toward
increasing polar angle. We denote the proton and hyperon transverse
polarization vectors by $S_T$ and $S_{\Lambda T}$, respectively; their
magnitudes give the corresponding degrees of polarization.
We measure the proton-spin
azimuth $\phi_S$ in its transverse plane and the hadron azimuth
$\phi_H$ about the jet axis, using these axes. The hyperon-spin and
hadron-pair azimuths below use the corresponding outgoing transverse
axes. Table~\ref{tab:transfer-analyzers} summarizes the {transverse spin
analyzers and their azimuthal correlations} considered here.

\begin{table}[htbp]
\centering\small
\setlength{\tabcolsep}{4pt}
\renewcommand{\arraystretch}{1.15}
\begin{tabular}{@{}>{\raggedright\arraybackslash}p{.20\linewidth}
                  >{\raggedright\arraybackslash}p{.26\linewidth}
                  >{\raggedright\arraybackslash}p{.47\linewidth}@{}}
\toprule
Measurement & Selected correlation & Fragmentation and jet ingredients\\
\midrule
Polarized hyperon
 & $\cos(\phi_S-\phi_\Lambda)$ in the hyperon spin difference
 & Collinear transverse-spin fragmentation $H_1^{\Lambda/c}$\\[3pt]
Hadron in a jet
 & $\sin(\phi_S-\phi_H)$
 & $j_\perp{\cal H}_1^{\perp,h/c}/(z_hM_h)$; transverse jet matching
   and TMD fragmentation with in-jet recoil\\[3pt]
Ordered hadron pair in a jet
 & $\sin(\phi_S-\phi_R)$
 & $-|\boldsymbol R_T|H_1^{\sphericalangle,h_1h_2/c}/M_{12}$, convolved with
   dihadron jet matching\\[3pt]
One-point energy correlator
 & $\sin(\phi_S-\phi_h)$
 & Energy-weighted Collins fragmentation for selected species,
   with the corresponding jet matching\\[3pt]
Two-point energy correlator
 & $\sin(\phi_S-\phi_R)$
 & Energy-weighted interference fragmentation for selected pairs,
   with the corresponding jet matching\\
\bottomrule
\end{tabular}
\caption{{Transverse spin analyzers for the measurements} described by
Eqs.~\eqref{eq:pp-transfer-analyzer} and~\eqref{eq:ep-transfer-analyzer}.
The displayed correlations multiply scalar structure functions. Here
$M_h$ is the identified-hadron mass, $M_{12}$ is the pair invariant
mass, and $\boldsymbol R_T$ is the transverse relative momentum of the ordered
pair. The energy weights are defined in
Eq.~\eqref{eq:transfer-energy-weights}.}
\label{tab:transfer-analyzers}
\end{table}

\paragraph{Polarized hyperons.}
In $p^\uparrow p\to\Lambda^\uparrow X$ and
$\ell p^\uparrow\to\Lambda^\uparrow X$, the collinear fragmentation
function $H_1^{\Lambda/c}$ describes the correlation between the
transverse quark and hyperon spins. Its convolution with quark transversity
distributions in the proton and the production hard function gives a leading-twist
spin-transfer structure function~\cite{Kang:2023elg}.
Let $\phi_\Lambda$ be the hyperon-spin azimuth in its transverse
basis. In the full spin difference
$\dd\sigma(S_T,S_{\Lambda T})-\dd\sigma(S_T,-S_{\Lambda T})$,
the coefficient of $\cos(\phi_S-\phi_\Lambda)$ is
$|S_T||S_{\Lambda T}|F_T[H_1^{\Lambda/c}]$, with $F_T$ chosen for the
collision system. Projection onto this cosine selects
$\widehat\sigma_T=(\widehat\sigma_{T,XX}+\widehat\sigma_{T,NN})/2$.
Replacing the $\Lambda$ by another transversely polarized
spin-$\tfrac12$ hadron changes the fragmentation function while
retaining the same production hard functions.

\paragraph{Collins hadron-in-jet production.}
The Collins fragmentation function describes the correlation between
the transverse quark spin and the azimuth of an unpolarized hadron.
STAR has measured the
charged-pion Collins asymmetry in transversely polarized $pp$
collisions at RHIC~\cite{STAR:2017akg}. The phenomenological analysis
of Ref.~\cite{Kang:2017btw} used Born production hard factors; the
hard functions of Sec.~\ref{sec:pp-transfer} supply {the} inclusive NLO
production correction. In $\ell p^\uparrow\to\mathrm{jet}(h)+X$
with the scattered lepton unobserved, the hard function of
Sec.~\ref{sec:pheno} supplies the production correction to the
Collins asymmetry studied as a probe of transversity at the
EIC~\cite{DAlesio:2026pca,Flore:2026gtk}.

For either collision system, write
$\dd\mathcal P=\dd\eta_J\,\dd^2\boldsymbol p_T\,\dd z_h\,
\dd^2\boldsymbol j_\perp$ and define the measured structure functions
by
\begin{align}
 \frac{\dd\sigma(S_T)}{\dd\mathcal P}
 &=F_{UU}+|S_T|\sin(\phi_S-\phi_H)F_{UT}^{\rm C}
  +\text{other angular harmonics},\nonumber\\
 A_{UT}^{\rm C}&=\frac{F_{UT}^{\rm C}}{F_{UU}}.
 \label{eq:collins-structure}
\end{align}
Here $F_{UU}$ is the azimuth-independent spin-averaged structure
function, and $F_{UT}^{\rm C}$ is the coefficient of the Collins
modulation after removing the proton polarization. Both refer to the
same kinematics and cuts. With uniform angular acceptance, the
asymmetry is the moment
\begin{equation}
 A_{UT}^{\rm C}=\frac{2}{|S_T|}
 \frac{\int\dd\phi_S\dd\phi_H\,\sin(\phi_S-\phi_H)
 [\dd\sigma(S_T)-\dd\sigma(-S_T)]}
 {\int\dd\phi_S\dd\phi_H\,[\dd\sigma(S_T)+\dd\sigma(-S_T)]}.
 \label{eq:collins-moment}
\end{equation}
The remaining measured variables are held fixed in these angular
integrals. This convention agrees with Ref.~\cite{Kang:2017btw}.

With $\epsilon_T^{12}=+1$, the spin-dependent Collins fragmentation
correlator contains the transverse vector
$-\epsilon_T^{ij}j_{\perp j}$. The diagonal components of the finite
production spin tensor are $\widehat\sigma_{T,XX}$ and
$\widehat\sigma_{T,NN}$. Contracting this tensor with the incoming
spin and the Collins vector gives the angular dependence
\begin{equation}
\begin{aligned}
 &-\widehat\sigma_{T,XX}\cos\phi_S\sin\phi_H
  +\widehat\sigma_{T,NN}\sin\phi_S\cos\phi_H\\[3pt]
 &=\frac{\widehat\sigma_{T,XX}+\widehat\sigma_{T,NN}}2
     \sin(\phi_S-\phi_H)
   +\frac{\widehat\sigma_{T,NN}-\widehat\sigma_{T,XX}}2
     \sin(\phi_S+\phi_H).
\end{aligned}
 \label{eq:collins-tensor}
\end{equation}
The angular moment in Eq.~\eqref{eq:collins-moment} therefore selects
$\widehat\sigma_T$. The {Collins transverse spin analyzer} is
${\cal A}_{\rm C}^c=j_\perp{\cal H}_1^{\perp,h/c}/(z_hM_h)$.
In lepton--proton collisions the structure functions are
\begin{align}
 F_{UU}^{\ell p}
 &=\frac{1}{S}\sum_{a,c}
 f_1^{a/p}\otimes{\cal K}_{ac}^{U}\otimes{\cal D}_1^{h/c},
 \nonumber\\
 F_{UT}^{{\rm C},\ell p}
 &=\frac{1}{S}\sum_{a,c=q,\bar q}
 h_1^{a/p}\otimes{\cal K}_{ac}^{T}
 \otimes\frac{j_\perp}{z_hM_h}{\cal H}_1^{\perp,h/c}.
 \label{eq:collins-factorization}
\end{align}
The unpolarized sum includes quark, antiquark, and gluon channels.
The $pp$ Collins structure follows from
Eq.~\eqref{eq:pp-transfer-analyzer} with the same ${\cal A}_{\rm C}^c$;
its spin-averaged counterpart uses the unpolarized production
hard functions and ${\cal D}_1^{h/c}$.

\paragraph{Dihadron interference fragmentation.}
Two unpolarized hadrons provide a complementary probe of transversity
through the correlation between the azimuth of the pair's relative
transverse momentum and the proton spin~\cite{Bacchetta:2004it}. Fix the hadron ordering and define
$R=(P_1-P_2)/2$.
{In the collision center-of-mass frame, $\boldsymbol R_T$ is the
component of $\boldsymbol R$ perpendicular to
$\boldsymbol P_1+\boldsymbol P_2$, and $\phi_R$ is its azimuth in the
corresponding outgoing transverse basis.} {After integration} over the
pair's total transverse momentum relative to the fragmenting quark,
the spin dependence is described by the collinear interference
fragmentation function $H_1^{\sphericalangle,h_1h_2/c}$ at fixed pair
mass $M_{12}\ll p_T$.

In the convention of Ref.~\cite{Kang:2026qbm}, the coefficient of
$\sin(\phi_S-\phi_R)$ contains the {transverse spin analyzer}
\[
 -\frac{|\boldsymbol R_T|}{M_{12}}H_1^{\sphericalangle,h_1h_2/c}.
\]
The minus sign is defined relative to the Collins convention in
Eq.~\eqref{eq:collins-tensor}, with the hadron ordering fixed as
above. For a pair identified inside a jet with $M_{12}\ll p_T R$,
${\cal A}^c$ is the dihadron fragmenting jet function matched onto
this collinear {transverse spin analyzer}. Equations~\eqref{eq:pp-transfer-analyzer}
and~\eqref{eq:ep-transfer-analyzer} then describe the selected
coefficient of $\sin(\phi_S-\phi_R)$ in both collision systems.

\FloatBarrier
\subsubsection{Energy-weighted spin correlations}

One- and two-point energy correlators measure weighted versions of
the single-hadron and hadron-pair correlations. The one-point
observable proposed in Ref.~\cite{Gao:2025evv} weights the
single-hadron distribution by the hadron's energy fraction at fixed angle
$\theta$ {to the jet axis}, with $\theta\ll R$.
{Energy-weighted dihadron fragmentation functions describe near-side
two-point correlators~\cite{Kang:2025zto}. Their chiral-odd counterparts
couple to transversity in spin-dependent
correlators~\cite{Kang:2026pro}.} The two-point observable
of Ref.~\cite{Kang:2026qbm} weights pairs at fixed opening angle
$\chi$ and measures the azimuth of the pair's relative transverse
momentum in the near-side, small-opening-angle regime. The event weights are
\begin{align}
 {\cal W}_1(\theta,\phi)
 &=\sum_{h\in J}z_h\,
 \delta(\theta-\theta_h)\delta(\phi-\phi_h),\nonumber\\
 {\cal W}_2(\chi,\phi)
 &=\sum_{(h_1,h_2)\in J}z_1z_2\,
 \delta(\chi-\chi_{12})\delta(\phi-\phi_R).
 \label{eq:transfer-energy-weights}
\end{align}
At the retained small-angle accuracy, the jet light-cone fractions
$z_h$ and $z_i$ are the corresponding energy fractions. The sums
select the measured hadron species; the second sum contains distinct
hadrons, counts each selected pair once, and uses the ordering fixed
in the definition of $R$. Weighting and summing the Collins and
interference fragmentation contributions gives the corresponding
{transverse spin analyzers} ${\cal A}^c$. These are convolved with the production
hard functions in
Eqs.~\eqref{eq:pp-transfer-analyzer}
and~\eqref{eq:ep-transfer-analyzer}.

STAR has measured the $\sin(\phi_S-\phi_h)$ and
$\sin(\phi_S-\phi_{hh})$ modulations of one- and two-point correlators
for charged pions and pion pairs in transversely polarized $pp$
collisions at RHIC~\cite{STARCollaboration:2026kaa}.
{For the two-point measurement we choose $(h_1,h_2)=(\pi^+,\pi^-)$,
so that the relative direction runs from $\pi^-$ to $\pi^+$.
The experimental angle $\phi_{hh}$ is measured along this direction
in the plane transverse to the jet. With the pair and jet directions
identified in the leading-power collinear matching,
$\phi_{hh}=\phi_R$ in the outgoing $(X,N)$
basis~\cite{Kang:2026qbm}.} These measurements
probe energy-weighted chiral-odd fragmentation in conjunction with
quark transversity distributions in the proton. The new $pp$ hard functions provide the NLO
inclusive production term for these correlations. {The
{direct-lepton} hard function supplies the corresponding production term for
hadron and hadron-pair correlations in single-inclusive jets at the EIC,
including their energy-weighted forms, with the scattered lepton
integrated over.}

\section{Discussion and outlook}
\label{sec:outlook}

{We have presented \sysname, a workflow in which AI agents develop and
execute process-specific programs for analytic perturbative QCD
calculations, where physicists specify the processes, observables, spin
projections and conventions. At NLO, we have validated this workflow
against unpolarized and polarized collinear hard functions, as well as
TMD matching coefficients and soft functions. Pole cancellation and RG
evolution provide further checks. These applications demonstrate the
reuse of shared infrastructure across processes and spin projections.}

{Using the \sysname \ workflow, we have calculated the new analytic NLO
hard functions for single-inclusive transverse spin transfer in
proton--proton collisions and in lepton--proton collisions with the
scattered lepton unobserved. The proton--proton results cover all
partonic channels contributing to transverse spin transfer. These hard
functions enter the factorized spin-dependent cross sections through
convolutions with the incoming PDFs and the fragmentation and jet
functions for the measured final state. The resulting cross sections
can be used to extract quark transversity distributions in the proton
from hadron polarization and azimuthal correlations in hadron and jet measurements at RHIC and the
EIC. The analytic hard functions for each partonic channel improve
numerical control of cancellations among polarized contributions and
can be reused in repeated convolutions as the fit inputs change.}

{The accompanying coefficient tables and calculation programs make
the results reproducible and provide reusable components for further
calculations. Building on the programs developed here, we plan to
extend \sysname\ to analytic multi-loop QCD calculations and present
these extensions in future publications.}

\acknowledgments
Z.K. and C.Z. are supported by the National Science Foundation under grant
No.~PHY-2515057. W.V. is supported by Deutsche Forschungsgemeinschaft (DFG)
through the Research Unit FOR 2926 (Project No.~409651613).

\clearpage
\appendix

\section{Analytic coefficients and ancillary conventions}
\label{app:coefficients}
\label{app:transverse-coefficients}

The analytic coefficient tables accompanying \sysname\ are provided
with the calculation programs at
\url{https://github.com/qcdphysics/FeynFacet}. For single-inclusive
production, the NLO coefficient lists are ordered as $\{R,B,C,A\}$,
with a separate prefactor $\mathcal N$. Writing the normalized Born
hard coefficient as
$\widehat{\sigma}_{\mathrm B}(s,v)$, the hard function through NLO is
\begin{align*}
 \widehat{\sigma}(s,v,w)
 ={}&\widehat{\sigma}_{\mathrm B}(s,v)\,\delta(1-w)
 +\mathcal N(s,v,w)\,R(v,w)\\
 &+\mathcal N(s,v,1)
 \left[B(v)\cD_0(w)+C(v)\cD_1(w)
       +A(v)\delta(1-w)\right].
\end{align*}
The Born coefficient includes its couplings and kinematic normalization.
The NLO prefactor is evaluated at $w=1$ for the endpoint distributions,
as in Eq.~\eqref{eq:distribution-form}.

The hadronic tables retain general SU($N_c$) color factors, $N_f$,
and independent renormalization, initial-state factorization and
fragmentation scales. Their NLO prefactor is
$\mathcal N_{pp}=\alpha_s^3/(\pi s v w)$, and their Born coefficients
include $\alpha_s^2/s$. The incoming-TT coefficients use the
$\cos(2\Phi)$ projection of Sec.~\ref{sec:pp-tt}.
The file \texttt{Coefficients/ProtonProtonTransfer.wl} contains the
seven transfer classes, with separate \texttt{XX} and \texttt{NN}
coefficient lists and Born terms in the convention of
Eq.~\eqref{eq:pp-transfer-finite}. The scale symbols
\texttt{LR}, \texttt{LF} and \texttt{LFF} denote $L_R$, $L_F$ and
$L_F'$, respectively, with a common factorization scale for the two
incoming partons.

For lepton transfer, \texttt{Coefficients/Collins.wl} uses
$\mathcal N_\ell=\alpha_{\rm em}^2\alpha_s C_F e_q^2/(\pi s v w)$
and the physical normalization of
Eq.~\eqref{eq:transverse-finite-hard}. Its NLO list is
$\{R_T,B_T,C_T,A_T\}$, and \texttt{LI}, \texttt{LO} and \texttt{LL}
denote $L_F$, $L_F'$ and $L_L$. The \texttt{ResolvedBorn} table in
the same file contains the $D$-dimensional spin-projected squared
matrix elements $|\mathcal M|^2/g_s^4$. Its \texttt{Transfer}
entries give the \texttt{NN}, \texttt{XX} and \texttt{Collins}
projections, with the last equal to the average of the first two.
The four-dimensional \texttt{Collins} entries are the reduced squared matrix elements
$\mathcal T_{ab\to c}$ in Eq.~\eqref{eq:resolved-transverse-born}.
The proton parton is listed first, and
$t=(p_a-p_c)^2$, where $p_a$ and $p_c$ are the momenta of that
incoming parton and the tagged outgoing parton. The
resolved-photon Born hard coefficient is
\[
 \widehat{\sigma}_{\mathrm B}(s,v)
 =\frac{\alpha_s^2}{s v}\,
 \mathcal T_{ab\to c}\bigl(s,-s(1-v),-s v\bigr).
\]

The file \texttt{Coefficients/TMD.wl} contains the one-loop
contributions $a_s\cI^{(1)}_{ij}$, $a_s\cC^{(1)}_{ij}$ and
$a_sS_i^{(1)}$ in its \texttt{Beam}, \texttt{TMDPDF} and
\texttt{Soft} entries, respectively, with
$a_s=\alpha_s/(4\pi)$. The variable \texttt{x} denotes the matching
fraction $\xi$; \texttt{Lp}, \texttt{Lv}, \texttt{LQ} and
\texttt{Lh} denote $L_\perp$, $L_\nu$, $L_Q$ and
$\ln(\zeta/\mu^2)$. The external-momentum logarithm $L_Q$ is held
fixed in the convolution over $\xi$.

\Needspace{22\baselineskip}
\paragraph{Lepton--quark transverse spin transfer.}

For $\ell q^\uparrow\to q^\uparrow+X$, the lepton is unpolarized
and the scattered lepton is unobserved. The endpoint coefficients
are given in Eq.~\eqref{eq:transverse-endpoints}. In the normalization
of Eq.~\eqref{eq:transverse-finite-hard}, define
$\rho=1-v+vw$ and $L_L=\ln(\mu_L^2/s)$, where $\mu_L$ is the
photon-in-lepton factorization scale. The complete regular
coefficient is
\begin{align}
 R_T(v,w)=\frac{2v}{(1-v)^2}\Bigg\{&
 \frac{w(1-v)^2}{1-vw}+2(w+vw^2)L_F'
 -\frac{w\,p_L}{1-vw}L_L \notag\\
 &+w(3+v^2)\ln w
 +\frac{\rho(1-v)^2}{v}\ln \rho
 +\frac{\mathcal P_a}{v}\ln(1-v)\notag\\
 &-\frac{\mathcal P_b}{v}\ln(1-vw)
 +\frac{\mathcal P_c}{1-vw}\ln v
 +\frac{\mathcal P_d}{1-vw}\ln(1-w)\Bigg\},
 \label{eq:transverse-regular}
\end{align}
where the auxiliary polynomials are
\begin{align}
 p_L={}&1+v^2-2vw-2v^2w+2v^2w^2,\notag\\
 \mathcal P_a={}&-1+3v-3v^2+v^3-2vw+2v^2w-4v^2w^2,\notag\\
 \mathcal P_b={}&-1+2v-v^2+vw+2v^2w+v^3w-4v^2w^2,\notag\\
 \mathcal P_c={}&-w+v^2w-2vw^2-2v^2w^2+4v^2w^3,\notag\\
 \mathcal P_d={}&1-2v+v^2-4w-vw+2v^2w-v^3w
          +vw^2-2v^2w^2+v^3w^2+4v^2w^3.
\end{align}
All logarithms have positive arguments for $0<v,w<1$.
At fixed $0<v<1$, the $\ln(1-w)$ singularity is integrable;
the nonintegrable threshold terms are contained in $\cD_0$
and $\cD_1$.

Equation~\eqref{eq:transverse-regular} uses the endpoint-prefactor
convention specified above. If the full $1/w$ prefactor multiplies
the plus distributions, the corresponding regular coefficient is
\[
 \widetilde R_T(v,w)
 =R_T(v,w)-B_T(v)-C_T(v)\ln(1-w).
\]
The equivalence follows from
\[
 \frac{1}{w}\cD_0(w)=\cD_0(w)+\frac{1}{w},
 \qquad
 \frac{1}{w}\cD_1(w)=\cD_1(w)+\frac{\ln(1-w)}{w}.
\]

\section{Wilson-line amplitudes and TMD master integrals}
\label{app:tmd-details}

\subsection{Wilson-line operators in \software{FeynArts}}
\label{sec:tmd-model}

We implement the operators in Eqs.~\eqref{eq:tmd-bilocal-operators}
and~\eqref{eq:tmd-soft-cut-operator} in a \software{FeynArts} model
that extends massless QCD
with directed fundamental and adjoint Wilson lines along $n$ and
$\bar n$. For a line with tangent $v$, the gluon emission vertex
is proportional to $v^\mu$, and the eikonal propagator contains
the routed denominator $v\cdot q$. The operator definition fixes
its causal prescription; the conjugate amplitude carries the
complex-conjugate prescription.

Nondynamical external fields mark the quark, gluon and soft operator
insertions. Auxiliary couplings identify these insertions during
diagram generation and are divided out after squaring. The quark
insertion leaves a Dirac index open for the unpolarized projection.
The gluon insertion includes the one-gluon term and the two-gluon
contact from the commutator in $n_\nu F^{\nu\mu}$, together with the
adjoint Wilson-line couplings. The soft insertion connects the
Wilson lines along the two lightlike directions.

The quark and gluon operators each generate one Born diagram.
The real-emission amplitudes contain two diagrams in each of the
$qq$ and $qg$ beam channels, one in $gq$, and four in $gg$.
In the $gg$ channel, the three-gluon vertex, field-strength contact
and Wilson-line attachments contribute to the same current.
For each soft representation, one diagram describes emission from
each Wilson line. The summed currents satisfy the Ward identities:
replacing an external-gluon polarization by its momentum gives zero
when the other external gluons are physical. We then sew the
amplitudes to their conjugates and perform the spin and color
contractions. The Born matrix elements reproduce the normalization
$B^{(0)}_{i/j}=\delta_{ij}\delta(1-\xi)$.

\subsection{Measured master integrals}
\label{app:tmd-integrals}

With the Fourier measurement and exponential regulator of
Sec.~\ref{sec:tmd-setup}, define the normalized transverse integral
\begin{equation}
 \mathcal T[f]
 =4\pi\mu^{2\epsilon}
 \left(\frac{e^{\gamma_E}}{4\pi}\right)^\epsilon
 \int\frac{\dd^{\,2-2\epsilon}\boldsymbol k_T}
 {(2\pi)^{2-2\epsilon}}\,
 e^{i\boldsymbol b_T\cdot\boldsymbol k_T}
 \frac{f(k_T)}{k_T^2},
\end{equation}
where $k_T=|\boldsymbol k_T|$. The rapidity regulator is removed
by retaining the leading small-$\tau$ terms at fixed $\epsilon$,
with $\nu=1/\tau$, before taking the Laurent expansion.

Away from the longitudinal endpoint, the regulator drops out and
the transverse master is
\begin{equation}
 F_\perp(\epsilon,\Lp)
 =\mathcal T[1]
 =e^{-\gamma_E\epsilon}\Gamma(-\epsilon)e^{\epsilon\Lp}
 =-\frac1\epsilon-\Lp
 -\frac{\epsilon}{2}\left(\Lp^2+\zeta_2\right)
 +\order{\epsilon^2}.
 \label{eq:tmd-transverse-master}
\end{equation}
Its Euler representation involves
\begin{equation}
 T_0=\int_0^\infty\dd u\,\frac{u^{-1-\epsilon}}{1+u},
 \qquad
 T_1=\int_0^\infty\dd u\,\frac{u^{-\epsilon}}{(1+u)^2}.
\end{equation}
Integrating the derivative of $u^{-\epsilon}/(1+u)$ gives
$-\epsilon T_0-T_1=0$ for
$-1<\operatorname{Re}\epsilon<0${. The transverse master is}
\[
 F_\perp
 =\frac{e^{-\gamma_E\epsilon+\epsilon\Lp}}
 {\Gamma(1+\epsilon)}\,T_0.
\]

For soft emission, the regulated longitudinal integral gives
$2K_0(2\tau e^{-\gamma_E}k_T)$, so the measured soft master is
\[
 \mathcal M_{\rm soft}
 =\mathcal T\!\left[2K_0(2\tau e^{-\gamma_E}k_T)\right].
\]
Introduce
\[
 z=\frac{\bT^2}{b_0^2\tau^2},
 \qquad c=1+z>1,
 \qquad L_t=\ln(\mu^2\tau^2),
\]
with $b_0=2e^{-\gamma_E}$ and $\Lp-\ln z=L_t$.
The corresponding Euler integrals are
\begin{align}
 E_0&=\int_0^\infty\dd u\,
       \frac{u^{-1-\epsilon}(1+cu)^\epsilon}{1+u},\notag\\
 E_1&=\int_0^\infty\dd u\,
       \frac{u^{-\epsilon}(1+cu)^{\epsilon-1}}{1+u},
 \qquad
 E_2=\int_0^\infty\dd u\,
       \frac{u^{-\epsilon}(1+cu)^\epsilon}{(1+u)^2}.
\end{align}
Integrating the derivative of
$u^{-\epsilon}(1+cu)^\epsilon/(1+u)$ yields
\begin{equation}
 -\epsilon E_0+\epsilon cE_1-E_2=0,
 \label{eq:tmd-euler-ibp}
\end{equation}
with vanishing surface terms in the same strip of $\epsilon$.
The normalized soft master is
\begin{equation}
 \mathcal M_{\rm soft}
 =F_\perp z^{-\epsilon}E_0
 =-\frac{e^{-\gamma_E\epsilon+\epsilon L_t}
 \Gamma(1-\epsilon)}{\epsilon}\,E_0.
\end{equation}

The reductions $T_0=-T_1/\epsilon$ and
$E_0=cE_1-E_2/\epsilon$ leave integrals that are finite at
$\epsilon=0$. We implement these identities with \software{Kira}
and evaluate $T_1$, $E_1$ and $E_2$ with \software{SubTropica}.
The expansions of $T_1$ and $E_2$ through $\epsilon^2$ and of
$E_1$ through $\epsilon$ determine $F_\perp$ and $E_0$ through
$\epsilon$, as required for the finite term of their product.
A numerical comparison of the Bessel--Fourier and Euler
representations checks the normalization of $\mathcal M_{\rm soft}$.

At leading power in $\tau$, the soft master has the compact form
\begin{equation}
 \mathcal M_{\rm soft}
 =F_\perp(\epsilon,\Lp)
 \left[\Lp+\Lnu-\gamma_E-\psi(-\epsilon)\right],
\end{equation}
where $\Lnu=\ln(\nu^2/\mu^2)$ and $\psi$ is the digamma function.
Its Laurent expansion is
\begin{equation}
 \mathcal M_{\rm soft}
 =\frac1{\epsilon^2}-\frac{\Lnu}{\epsilon}
 -\frac{\Lp^2}{2}-\Lp\Lnu-\frac{\zeta_2}{2}
 +\order{\epsilon}.
\end{equation}
In the convention of Eq.~\eqref{eq:tmd-soft-expansion}, the bare
coefficient is $S_i^{(1),\mathrm{bare}}=4C_i\mathcal M_{\rm soft}$,
with $C_q=C_F$ and $C_g=C_A$. Its poles and finite term reproduce
Eq.~\eqref{eq:tmd-soft-poles}; subtracting the poles gives the
renormalized soft coefficient in Eq.~\eqref{eq:tmd-soft}.

The same master also determines the regulated longitudinal endpoint
of the diagonal beam channels. The singular real-emission term
contributes
\begin{equation}
 4a_sC_i\left\{
 F_\perp\,\cD_0(\xi)
 +\delta(1-\xi)
 \left[\mathcal M_{\rm soft}+\frac{L_Q}{2}F_\perp\right]
 \right\}
\end{equation}
to the bare matrix element, where $\cD_0(\xi)=[1/(1-\xi)]_+$ and
$L_Q=2\ln(Q/\nu)$, with $Q=xP^+$ as defined in
Sec.~\ref{sec:tmd-setup}.
The soft-overlap subtraction removes
$a_s\delta(1-\xi)S_i^{(1),\mathrm{bare}}$, cancelling the
$\mathcal M_{\rm soft}$ term. Adding the regular real-emission
terms and the operator and PDF counterterms gives the beam
matching coefficients of Sec.~\ref{sec:tmd-beam}.

\section{Checks of amplitudes and hard functions}
\label{app:spec}
\label{app:acceptance}

\begin{figure}[t]
 \centering
 \includegraphics[width=\textwidth]{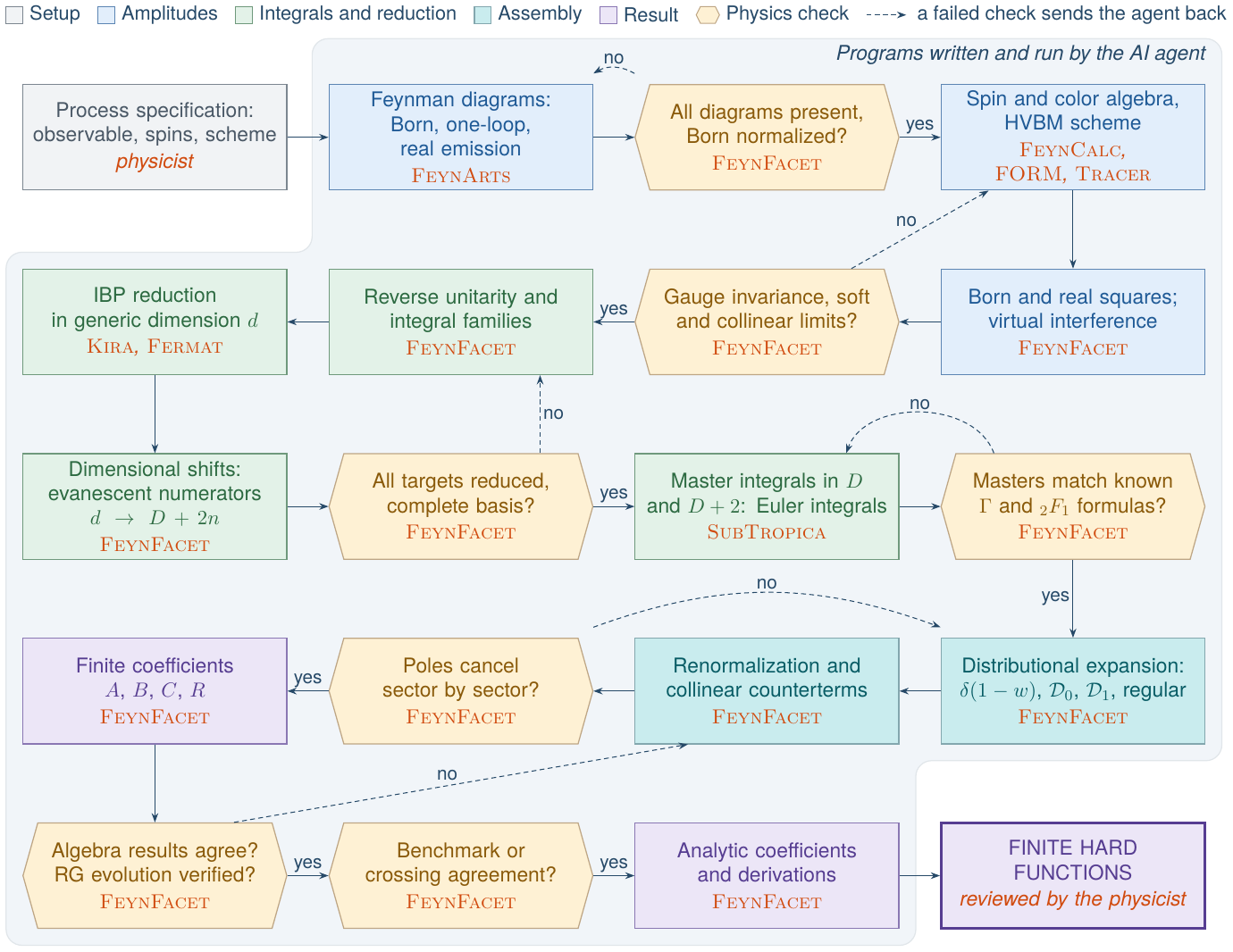}
 \caption{Stages and checks of the \sysname\ workflow summarized in
 Figure~\ref{fig:workflow}. Rectangles denote calculation stages, with
 package and program names in small capitals; hexagons denote physics
 checks. Dashed arrows indicate the return to an earlier stage when a
 check fails. The shaded region contains the programs developed and
 executed by the agents. Physicists specify the process, observable,
 spin projection and conventions, and review the finite hard functions.}
 \label{fig:workflow-chart}
\end{figure}

Comparisons between the \software{FeynCalc}, \software{FORM} and
\software{Tracer} implementations test independent evaluations of the
spin and color algebra. The common reduction, integration and assembly
programs are tested through comparisons of master integrals and finite
hard functions. {These programs are applied
separately to each algebra result for all five incoming-TT classes and
all seven transfer-TT classes. The finite coefficients from the three
routes agree at six kinematic points.} Table~\ref{tab:validation} summarizes the coverage of
these comparisons. Figure~\ref{fig:workflow-chart} locates the checks
within \sysname\ and shows how failed checks guide revisions of the
calculation programs.

\paragraph{Amplitudes, gauge invariance and permutation symmetry.}
The algebra results are compared after aligning spinor orientations,
momentum signs, dimensional prescriptions and normalization factors.
The comparisons include the Born hard functions entering the
collinear counterterms.

For the TMD operators, the summed emission currents satisfy the Ward
identities when the other external gluons are physical, as described
in Appendix~\ref{sec:tmd-model}. For each hadronic class, the
spin-averaged squared matrix element of the first real subprocess gives
the same result with different physical-gluon reference vectors at
three exact-rational choices of kinematics, color and dimension.
Exchanging the two unobserved gluons in $gg\to ggg$ and $qg\to qgg$
also gives identical squared matrix elements in UU and LL at three
exact-rational points with $D=4$. For quark-tagged $qg\to qgg$, this
check separately reconstructs the interference sums. For $gg\to ggg$,
the channel definition is also checked to include the symmetry factor
$1/2!$ for the two identical unobserved gluons.

\paragraph{Scalar master integrals.}
In the single-inclusive real-emission calculation, the two cut
propagators represent the unobserved two-particle phase space.
For the lepton--quark calculation, the ordinary and {dimensionally shifted}
phase-space master expansions agree with their Gamma-function
expressions through $\epsilon^2$ and $\epsilon$, respectively.
The bubble-master expansion agrees with its Gamma-function expression
through $\epsilon$.

For the hypergeometric functions entering the ordinary and
{dimensionally shifted} angular masters, numerical comparisons at fixed
kinematics show the expected $\epsilon^3$ scaling of the difference
between the exact functions and their truncated expansions.
Direct numerical evaluations of the angular integrals likewise agree
with the analytic expansions to the expected orders in $\epsilon$.
These comparisons test both ordinary masters and the {dimensionally shifted}
masters generated by evanescent numerators. The corresponding checks
of the measured TMD integrals are described in
Appendix~\ref{app:tmd-integrals}.

\paragraph{Unresolved limits, pole cancellation and {RG evolution}.}
At $D=4$, the $qq'\to qq'g$ squared matrix elements reproduce the
soft-gluon limit and the collinear limit on the incoming $q$ leg
at general kinematic ratios in UU, LL and the $XX$ and $NN$
components of transfer TT.
After assembly, the double and single poles cancel separately in the
$\delta(1-w)$, $\cD_0$ and regular sectors of the hard functions.

For each of the three direct-lepton channels in UU and LL, we
calculate the collinear pole residues associated with the incoming
parton, fragmentation and photon emission from the lepton. Their sum
reproduces the total collinear pole residue. {We check the RG
evolution of the finite hard function against these residues.} With
$\mu_R=\mu_F=\mu_F'=\mu_L=\mu$, differentiating the finite hard
function with respect to $\ln\mu^2$ gives minus this total residue.
{For transverse spin transfer in lepton--proton collisions, we
check the RG evolution over the three factorization scales
independently. The RG evolution of the hard function cancels that of
the} {incoming-quark PDF, the quark FF}
and the photon distribution in the lepton. These terms are calculated by
convolving the Born hard functions with their respective leading-order
splitting kernels. Equation~\eqref{eq:lepton-photon-scale-check}
shows the photon-scale cancellation explicitly.

\paragraph{Finite transverse-spin-transfer hard functions.}
\leavevmode{The comparisons of dimensionally regulated expressions in}
Sec.~\ref{sec:pp-transfer-crossing} test the crossing relations between
the incoming TT and transfer TT channels in proton--proton
collisions. The subsequent direct,
crossed and independent finite assemblies use common $D$-dimensional
Born tensors, reduced integrals and master-integral values.
The independent implementation performs the endpoint expansion,
coupling renormalization and collinear subtraction. At symbolic
kinematics and scales, it reproduces the regular, $\cD_0$, $\cD_1$
and delta coefficients for both $XX$ and $NN$ in all seven transfer
classes, and verifies the equality of their finite $XX$ and $NN$
hard functions. The direct and crossed assemblies also give identical
finite transfer coefficients. Appendix~\ref{sec:caught} discusses the
dimensional terms whose omission changes these finite results.

\section{Algebraic complexity and {physical checks}}
\label{app:execution}

\subsection{Size of the real-emission calculation}
\label{sec:resources}

Table~\ref{tab:calculation-size} illustrates the size of the
real-emission algebra encountered in \sysname. For $n$ tree diagrams,
the {sum over the upper triangle} contains $n(n+1)/2$ products per spin
configuration, including diagonal terms and vanishing interferences.
Each spin projection requires its own spin contractions.

\begin{table}[htbp]
\centering
\small
\begin{tabular}{cccc}
\toprule
Real-emission sector & Observed parton & Diagrams & {Pairs per spin configuration}\\
\midrule
$\ell q\to\ell gq$ & $g$ & 2 & 3\\
$\ell g\to\ell q\bar q$ & $q$ & 2 & 3\\
$q_iq_k\to q_iq_k g$ ($k\ne i$) & $q_i$ & 5 & 15\\
$qg\to gqg$ & $g$ & 16 & 136\\
$gg\to q_i\bar q_i g$ & $q_i$ & 16 & 136\\
$gg\to ggg$ & $g$ & 25 & 325\\
$gg\to gq_i\bar q_i$ & $g$ & 16 & 136\\
\bottomrule
\end{tabular}
\caption{Representative real-emission calculation sizes. Each row
specifies an individual real-emission sector with a fixed tagged
parton. Counts apply to each of UU and LL in each algebra
implementation, before inclusive flavor multiplicities and
identical-particle weights.}
\label{tab:calculation-size}
\end{table}

The gluon-tagged and quark-tagged $gg\to gq_i\bar q_i$ sectors use
the same incoming-gluon LL projector. The choice of tagged parton
fixes a different observed momentum and changes which final-state
momenta are integrated over. Summing over the unobserved quark flavors
in the gluon-tagged sector gives a factor $N_f$, whereas a specified
tagged quark flavor carries unit weight.
For $gg\to g+X$, the $gg\to ggg$ and $gg\to gq_i\bar q_i$ real
contributions are combined with the gluon, ghost and closed-quark-loop
virtual contributions.

\FloatBarrier
\subsection{Algebraic and physical {checks}}
\label{sec:caught}

Comparing the spin and color results from the three algebra programs
allows discrepancies to be traced to color contractions, dimensional
spin terms or external-state sums before scalar integration. The
finite-coefficient comparisons then test the consequences of these
choices after renormalization and collinear subtraction.

\paragraph{Color contractions.}
Multi-gluon amplitudes contain networks of structure constants.
The \software{FeynCalc} implementation reduces these networks by
expressing each structure constant in terms of fundamental generators,
\begin{equation}
 f^{abc}
 =-2i\left[
 \operatorname{tr}(T^aT^bT^c)
 -\operatorname{tr}(T^bT^aT^c)
 \right],
 \qquad
 \operatorname{tr}(T^aT^b)=\frac12\delta^{ab},
\end{equation}
and then evaluating the traces and color contractions.
The \software{Tracer} implementation contracts fundamental generators
using Fierz identities, while \software{FORM} uses its color library.
The agreement between the algebra implementations summarized in
Table~\ref{tab:validation} tests these separate evaluations of the
color factors together with the independently evaluated spin
contractions.

\paragraph{Dimensional spin information.}
In annihilation channels with two helicity projectors on the same
fermion trace, the HVBM and fully anticommuting $\gamma_5$
prescriptions differ in their evanescent terms. The calculation
implements HVBM in \software{FORM} by expressing each helicity
insertion through an explicit four-dimensional Levi-Civita tensor
and retaining both insertions on the trace.

The dimension assigned to Lorentz indices also determines the
evanescent contractions. The \software{FORM} code sets
\texttt{Dimension D} before declaring the generic Lorentz indices,
so their contractions retain $D=4-2\epsilon$. Declaring these indices
in four dimensions would suppress the corresponding evanescent
terms. Comparisons of the spin tensors at symbolic $D$ test this
dimensional dependence as well as the four-dimensional limit.

Dimensional terms in the Born tensors and fragmentation phase space
contribute to finite hard functions when they multiply collinear
poles. The independent assembly described in
Appendix~\ref{app:acceptance} demonstrates this explicitly for
the $q^\uparrow q'\to q^\uparrow+X$ channel with $q'\ne q$.
Replacing its Born tensors by their
$D=4$ values before collinear subtraction changes the finite result
in both the $XX$ and $NN$ projections. A separate test removes the
dimensional factor
\begin{equation}
 (1-v+vw)^{2\epsilon}
 =1+2\epsilon\ln(1-v+vw)+\order{\epsilon^2}
\end{equation}
from the fragmentation counterterm while retaining the ordinary
Jacobian $vw/(1-v+vw)^2$. This change also alters the finite result
in each projection. In both tests, an $\order{\epsilon}$ term
multiplying a collinear $1/\epsilon$ pole accounts for the finite
contribution.

\paragraph{Physical external-gluon states.}
For an external gluon whose polarization is summed, the real-emission
calculation uses the symmetric physical projector
\begin{equation}
 P^{\mu\nu}(k,r)
 =-g^{\mu\nu}
 +\frac{k^\mu r^\nu+r^\mu k^\nu}{k\cdot r},
 \qquad
 k^2=r^2=0,
 \qquad k\cdot r\ne0,
\end{equation}
with a lightlike reference vector $r$. Incoming gluons in the LL
projection use the helicity-difference projector. In the spin-averaged
$qg\to qgg$ calculation, replacing all three external-gluon
polarization sums by $-g^{\mu\nu}$ includes unphysical states unless
their contributions are compensated.

Reference-vector independence tests the gauge consistency of the
external-state contractions. Exchanging identical unobserved gluons
tests the momentum assignments and the assembly of the real-emission
interferences. Appendix~\ref{app:acceptance} gives the coverage of
these checks.

\section{AI-assisted development and reproducibility}
\label{app:sessions}

In the \sysname\ calculations, physicists specified the processes,
observables, spin conventions and subtraction prescriptions, and
selected the calculational methods and literature benchmarks.
The agents reused the shared infrastructure described in
Sec.~\ref{sec:workflow} to develop and execute the programs for each
process. They compared the spin and color results from the three algebra
programs and used physical consistency checks to diagnose discrepancies
and revise the programs. Appendix~\ref{app:acceptance}
describes these comparisons, and Appendix~\ref{sec:caught} gives
{concrete examples of these algebraic and physical checks.}
{The step programs retain the checks used during development.}

The analytic coefficients can be used directly from the tables in
\path{Coefficients/*.wl}. These files contain plain \software{Wolfram Language}
expressions loaded with \texttt{Get}; their physical normalizations
are given in Appendix~\ref{app:coefficients}.
The file \path{Coefficients/README.md} describes {the coefficient formats and evaluation methods}, including the optional numerical interface
\path{Coefficients/evaluate.py}.

{Each directory \path{Processes/<name>/} contains numbered \texttt{.wl}
step programs for reproducing the derivation. The corresponding
\texttt{.nb} notebooks are generated from these programs.} Each step runs in a fresh \software{Mathematica} kernel,
reads the required outputs of earlier steps from \path{Results/},
and saves its results for subsequent steps. These programs evaluate
the amplitudes, scalar integrals and counterterms and run the
associated checks without further language-model calls.
For example, the lepton--quark calculation can be run from the
repository root with
\begin{quote}
\small
\texttt{python Checks/run\_process.py Processes/LeptonQuark --route FeynCalc}
\end{quote}
The \texttt{--route} option also selects \software{FORM} or
\software{Tracer}. Running \software{FeynCalc} first supplies the
saved expressions used in their algebra comparisons.
The runner executes the embedded checks, stops at a failed step,
and stores the kernel logs and per-step records under \path{Results/}.

The separate comparison programs and their settings are
described in \path{Checks/README.md} and
\path{Checks/validation-settings.wl}.
Dependency installation and configuration through \texttt{Paths.wl}
are described in \path{docs/INSTALL.md};
\texttt{CheckInstallation[]} verifies the tool paths.

Table~\ref{tab:agent-models} records the models and interfaces used
during development. The agents also assisted with documentation and
manuscript preparation.

\begin{table}[htbp]
\centering\small
\begin{tabular}{cc}
\toprule
Interface & Model identifier\\
\midrule
\software{OpenAI Codex} & \texttt{gpt-5.6-sol}\\
\software{OpenAI Codex} & \texttt{gpt-6-astra}\\
\software{OpenAI Codex} & \texttt{gpt-6.1-sol}\\
\software{OpenAI Codex} & \texttt{gpt-6-sol}\\
\software{Claude Code} & \texttt{claude-fable-5-1}\\
\software{Claude Code} & \texttt{claude-opus-5-5}\\
\bottomrule
\end{tabular}
\caption{Models and interfaces used in the \sysname\ development
sessions.}
\label{tab:agent-models}
\end{table}

The programs run in \software{Mathematica}~15.0.1. Diagram
generation uses \software{FeynArts}~3.12.
For algebra we use \software{FeynCalc}~10.2.1,
\software{FORM}/\allowbreak\software{TFORM}~5.0.1 and
\software{Tracer}~1.1.1. The \software{FORM} implementation uses
the official color package~\cite{vanRitbergen:1998pn}.
Integral reduction and integration use \software{Kira}~3.1 and
\software{SubTropica}~1.2.10 with \software{polymake}~4.4;
\software{Python}~3.12 is used for scripting.

\FloatBarrier

\bibliographystyle{JHEP}
\bibliography{references}

\end{document}